\documentclass[twocolumn]{aastex631}

\usepackage{savesym}
\savesymbol{tablenum}
\usepackage{siunitx}
\restoresymbol{SIX}{tablenum}

\usepackage[version=4]{mhchem}
\usepackage{graphicx,epsf}
\usepackage{subfigure}
\usepackage{graphicx}	
\usepackage{amsmath}	
\usepackage{amssymb}	
\usepackage{newtxtext,newtxmath}
\usepackage{siunitx}
\usepackage{footnote}

\newcommand{\Msun}{{\rm M}_\odot}
\newcommand{\Rsun}{{\rm R}_\odot}

\newcommand{\kms}{\textrm{km}\,\textrm{s}^{-1}}

\def\arcsec{\hbox{$^{\prime\prime}$}}

\def\sn{{SN~2024ggi}}
\newcommand{\mdot}{M$_{\odot}$~yr$^{-1}$}

\newcommand{\NSF}{NSF Graduate Research Fellow}

\DeclareRobustCommand{\ion}[2]{\relax\ifmmode\ifx\testbx\f@series{\mathbf{#1\,\mathsc{#2}}}\else{\mathrm{#1\,\mathsc{#2}}}\fi\else\textup{#1\,{\mdseries\textsc{#2}}}\fi}

\newcommand{\code}[1]{\texttt{#1}}
\def\heracles{{\code{HERACLES}}}
\def\cmfgen{{\code{CMFGEN}}}

\usepackage[scale=0.94]{tgbonum}
\usepackage[mode=text]{siunitx}

\makeatletter
\DeclareTextCompositeCommand{\r}{OT1}{A}{%
  \leavevmode\vbox{%
    \offinterlineskip
    \ialign{\hfil##\hfil\cr\char23\cr\noalign{\kern-1.15ex}A\cr}%
  }%
}
\makeatother

\shorttitle{Supernova 2024ggi}
\shortauthors{Jacobson-Gal\'an et al.}

\begin{document}

\title{SN~2024ggi in NGC~3621: Rising Ionization in a Nearby, CSM-Interacting Type II Supernova}

\correspondingauthor{Wynn Jacobson-Gal\'{a}n (he, him, his)}
\email{wynnjg@berkeley.edu}

\author[0000-0002-3934-2644]{W.~V.~Jacobson-Gal\'{a}n}
\affil{Department of Astronomy, University of California, Berkeley, CA 94720-3411, USA}
\affil{\NSF}

\author[0000-0002-5680-4660]{K.~W.~Davis}
\affil{Department of Astronomy and Astrophysics, University of California, Santa Cruz, CA 95064, USA}

\author[0000-0002-5740-7747]{C.~D.~Kilpatrick}
\affil{Center for Interdisciplinary Exploration and Research in Astrophysics (CIERA), Northwestern University, Evanston, IL 60202, USA}
\affiliation{Department of Physics and Astronomy, Northwestern University, Evanston, IL 60208, USA}

\author[0000-0003-0599-8407]{L.~Dessart}
\affil{Institut d’Astrophysique de Paris, CNRS-Sorbonne Université, 98 bis boulevard Arago, F-75014 Paris, France}

\author[0000-0003-4768-7586]{R.~Margutti}
\affil{Department of Astronomy, University of California, Berkeley, CA 94720-3411, USA}
\affil{Department of Physics, University of California, Berkeley, CA 94720, USA}

\author[0000-0002-7706-5668]{R.~Chornock}
\affil{Department of Astronomy, University of California, Berkeley, CA 94720-3411, USA}

\author[0000-0002-2445-5275]{R.~J.~Foley}
\affiliation{Department of Astronomy and Astrophysics, University of California, Santa Cruz, CA 95064, USA}

\author[0000-0002-6688-3307]{P.~Arunachalam}
\affiliation{Department of Astronomy and Astrophysics, University of California, Santa Cruz, CA 95064, USA}

\author[0000-0002-4449-9152]{K.~Auchettl}
\affil{Department of Astronomy and Astrophysics, University of California, Santa Cruz, CA 95064, USA}
\affil{School of Physics, The University of Melbourne, VIC 3010, Australia}

\author[0000-0003-4383-2969]{C.~R.~Bom}
\affiliation{Centro Brasileiro de Pesquisas F\'isicas, Rua Dr. Xavier Sigaud 150, 22290-180 Rio de Janeiro, RJ, Brazil}

\author[0000-0003-4553-4033]{R.~Cartier}
\affil{Instituto de Estudios Astrofísicos, Facultad de Ingenier\'ia y Ciencias, Universidad Diego Portales, Av. Ej\'ercito Libertador 441, Santiago, Chile}

\author[0000-0003-4263-2228]{D.~A.~Coulter}
\affil{Space Telescope Science Institute, Baltimore, MD 21218, USA}

\author[0000-0001-9494-179X]{G.~Dimitriadis}
\affil{School of Physics, Trinity College Dublin, The University of Dublin, Dublin 2, D02 PN40, Ireland}

\author[0000-0003-0913-4120]{D.~Dickinson}
\affil{Department of Physics and Astronomy, Purdue University, 525 Northwestern Avenue, West Lafayette, IN 47907, USA}

\author[0000-0001-7081-0082]{M.~R.~Drout}
\affil{David A. Dunlap Department of Astronomy and Astrophysics, University of Toronto, 50 St. George Street, Toronto, Ontario, M5S 3H4, Canada}

\author[0000-0003-4906-8447]{A.~T.~Gagliano}
\affil{The NSF AI Institute for Artificial Intelligence and Fundamental Interactions}

\author[0000-0002-8526-3963]{C.~Gall}
\affil{DARK, Niels Bohr Institute, University of Copenhagen, Jagtvej 128, 2200 Copenhagen, Denmark}

\author[0000-0001-6922-8319]{B.~Garretson}
\affil{Department of Physics and Astronomy, Purdue University, 525 Northwestern Avenue, West Lafayette, IN 47907, USA}

\author[0000-0001-9695-8472]{L.~Izzo}
\affil{DARK, Niels Bohr Institute, University of Copenhagen, Jagtvej 128, 2200 Copenhagen, Denmark}
\affil{Osservatorio Astronomico di Capodimonte, INAF, Salita Moiariello 16, Napoli, 80131, Italy}

\author[0000-0002-6230-0151]{D.~O.~Jones}
\affiliation{Institute for Astronomy, University of Hawai’i, 640 N. A’ohoku Pl., Hilo, HI 96720, USA}

\author[0000-0002-2249-0595]{N.~LeBaron}
\affil{Department of Astronomy, University of California, Berkeley, CA 94720-3411, USA}

\author[0000-0003-2736-5977]{H.-Y.~Miao}
\affil{Graduate Institute of Astronomy, National Central University, 300 Zhongda Road, Zhongli, Taoyuan 32001, Taiwan}

\author[0000-0002-0763-3885]{D.~Milisavljevic}
\affil{Department of Physics and Astronomy, Purdue University, 525 Northwestern Avenue, West Lafayette, IN 47907, USA}

\author[0000-0001-8415-6720]{Y.-C.~Pan}
\affil{Graduate Institute of Astronomy, National Central University, 300 Zhongda Road, Zhongli, Taoyuan 32001, Taiwan}

\author[0000-0002-4410-5387]{A.~Rest}
\affil{Space Telescope Science Institute, Baltimore, MD 21218}
\affiliation{Department of Physics and Astronomy, The Johns Hopkins University, Baltimore, MD 21218, USA}

\author[0000-0002-7559-315X]{C.~Rojas-Bravo}
\affiliation{Department of Astronomy and Astrophysics, University of California, Santa Cruz, CA 95064, USA}

\author[0000-0002-1420-3584]{A.~Santos}
\affil{Centro Brasileiro de Pesquisas F\'isicas, Rua Dr. Xavier Sigaud 150, 22290-180 Rio de Janeiro, RJ, Brazil}

\author[0000-0001-8023-4912]{H.~Sears}
\affil{Center for Interdisciplinary Exploration and Research in Astrophysics (CIERA), Northwestern University, Evanston, IL 60202, USA}
\affiliation{Department of Physics and Astronomy, Northwestern University, Evanston, IL 60208, USA}

\author[0000-0001-8073-8731]{B.~M.~Subrayan}
\affil{Department of Physics and Astronomy, Purdue University, 525 Northwestern Avenue, West Lafayette, IN 47907, USA}

\author[0000-0002-5748-4558]{K.~Taggart}
\affil{Department of Astronomy and Astrophysics, University of California, Santa Cruz, CA 95064, USA}

\author[0000-0002-1481-4676]{S.~Tinyanont}
\affil{National Astronomical Research Institute of Thailand, 260 Moo 4, Donkaew, Maerim, Chiang Mai, 50180, Thailand}

\begin{abstract}

We present UV/optical/NIR observations and modeling of supernova (SN) 2024ggi, a type II supernova (SN II) located in NGC~3621 at 7.2~Mpc. Early-time (``flash'') spectroscopy of \sn{} within +0.8~days of discovery shows emission lines of \ion{H}{i}, \ion{He}{i}, \ion{C}{iii}, and \ion{N}{iii} with a narrow core and broad, symmetric wings (i.e., IIn-like) arising from the photoionized, optically-thick, unshocked circumstellar material (CSM) that surrounded the progenitor star at shock breakout. By the next spectral epoch at +1.5~days, \sn{} showed a rise in ionization as emission lines of \ion{He}{ii}, \ion{C}{iv}, \ion{N}{iv/v} and \ion{O}{v} became visible. This phenomenon is temporally consistent with a blueward shift in the UV/optical colors, both likely the result of shock breakout in an extended, dense CSM. The IIn-like features in \sn{} persist on a timescale of $t_{\rm IIn} = 3.8 \pm 1.6$~days at which time a reduction in CSM density allows the detection of Doppler broadened features from the fastest SN material. \sn{} has peak UV/optical absolute magnitudes of $M_{\rm w2} = -18.7$~mag and $M_{\rm g} = -18.1$~mag that are consistent with the known population of CSM-interacting SNe~II. Comparison of \sn{} with a grid of radiation hydrodynamics and non-local thermodynamic equilibrium (nLTE) radiative-transfer simulations suggests a progenitor mass-loss rate of $\dot{M} = 10^{-2}~\Msun$~yr$^{-1}$ ($v_w = 50~\kms$), confined to a distance of $r < 5\times 10^{14}$~cm. Assuming a wind velocity of $v_w = 50~\kms$, the progenitor star underwent an enhanced mass-loss episode in the last $\sim 3$ years before explosion. 



\end{abstract}


\section{Introduction} \label{sec:intro}

Shock breakout (SBO) from a red supergiant (RSG) star is characterized by an optical depth of $\tau \approx c/v_{\rm sh}$, where $c$ is the speed of light and $v_{\rm sh}$ is the shock velocity. Consequently, the location and timescale of SBO photon escape is highly dependent on the density and extent of circumstellar material (CSM) that borders the RSG prior to explosion. In addition to light travel effects during SBO \citep{waxman17,Goldberg22}, the SBO signal can be significantly enhanced and elongated by the presence of high density CSM directly above the stellar surface \citep{Chevalier11, dessart17, haynie21}. Once the shock has ``broken out,'' the associated burst of high-energy radiation will ``flash ionize'' the surrounding medium -- observationally this manifests as a hot supernova (SN) continuum riddled with recombination lines from ionization CSM. However, to overcome the recombination timescale of the ``flash ionized CSM'' ($t_{\rm rec} \propto 1/n_e \approx $~hours-days for $n \approx 10^{7-10}$~cm$^{-3}$, $\rho \approx 10^{-14} - 10^{-17}$~g~cm$^{-3}$ at $r < 2R_{\star}$, where $R_{\star}$ is progenitor radius), SN ejecta interaction with dense CSM provides continuous photoionization of the medium for sufficiently large CSM densities (e.g., $\rho \gtrsim 10^{-14}$~g~cm$^{-3}$). 

For type II supernovae (SNe~II) interacting with such dense CSM, high-energy photons will be emitted from the shock front as the post-shock gas cools primarily through free-free emission ($T_{\rm sh} \approx 10^{5-8}$~K; \citealt{Chevalier12, Chevalier17}). This process then prolongs the formation of high-ionization recombination lines (e.g., \ion{He}{ii}, \ion{N}{iii/iv/v}, \ion{C}{iii/iv}) present during the ``flash ionization'' phase. Intriguingly, as recombination line photons try to exit the CSM, they electron-scatter off of free electrons in the ionized gas, which broadens the observed emission lines that will then appear as the combination of a narrow core and Lorentzian wings (i.e., ``IIn-like features''; \citealt{Chugai01, Dessart09, HuangES}). However, as the CSM density and optical depth to electron-scattering drops, these electron-scattering profiles will fade on a characteristic timescale ($t_{\rm IIn}$), with the SN photosphere then revealing the fastest moving SN material \citep{dessart17, Dessart23, wjg23, wjg24}.

Given the rapid evolution of CSM-interacting SNe~II in their first hours-to-days, ultra-rapid (``flash'') spectroscopy is essential to both capture the SBO signal and persistent photo-ionization of dense CSM, but also to constrain the composition and mass-loss history of the progenitor star in its final year(s) before collapse \citep{galyam14,Khazov16,yaron17,terreran22,wjg23}. To date, all-sky transient surveys have allowed for systematic discovery of SNe~II within days of first light. Through modeling of early-time SN~II spectra showing IIn-like features with non-LTE radiative transfer codes (e.g., \cmfgen; \citealt{hillier12, D15_2n}), numerous single-object studies indicate enhanced RSG mass-loss ($\dot{M} \approx 10^{-3}$-- $10^{-2}$~\mdot, $v_w\approx50$--100~km~s$^{-1}$) in the final years before explosion (e.g., PTF11iqb, \citealt{smith15}; SN~2013fs, \citealt{yaron17, dessart17}; SN~2014G, \citealt{Terreran16}; SN~2016bkv, \citealt{Hosseinzadeh18, Nakaoka18}; SN~2017ahn, \citealt{Tartaglia21}; SN~2018zd, \citealt{zhang20, Hiramatsu21}; SN~2019nyk, \citealt{Dastidar24}; SN~2020pni, \citealt{terreran22}; SN~2020tlf, \citealt{wjg22}; SN~2022jox, \citealt{Andrews23}; SN~2023ixf, \citealt{wjg23, Bostroem23, Teja23, Smith23, Zimmerman24}). Furthermore, sample studies suggest that $>$40\% of SNe~II discovered within 2 days of first light show IIn-like features from interaction with dense CSM \citep{bruch21, bruch23}. Additionally, relative to standard SNe~II, events with early-time IIn-like features are incredibly bright in the ultraviolet \citep{Irani23b, wjg24}. 

In this paper we present, analyze, and model photometric and spectroscopic observations of \sn{}, first discovered by the Asteroid Terrestrial-impact Last Alert System (ATLAS) \citep{Srivastav24, atlas24ggi, Chen24} on 2024-04-11 (MJD 60411.14). \sn{} was classified as a Type II SN \citep{Hoogendam24,Zhai24} and is located at $\alpha = 11^{\textrm{h}}18^{\textrm{m}}22.09^{\textrm{s}}$, $\delta = -32^{\circ}50'15.26^{\prime \prime}$ in host galaxy NGC~3621. We adopt a time of first light of MJD $60410.80 \pm 0.34$~days, which is based on the average between the times of last non-detection ($m_L = 19.5$~mag) and first detection $m_o = 18.9$~mag. To validate this estimate of first light, we first fit the ATLAS $o$-band light curve with a two-component power-law in the {\tt REDBACK} software package \citep{Sarin23} and derive a time of first light of MJD $60411.07 \pm 0.01$. Additionally, using a uniform prior distribution derived from the last ATLAS non-detection on MJD 60410.45, we fit the bolometric light curve of \sn{} to a suite of hydrodynamical models (e.g., see \citealt{Moriya23, Subrayan23}) which constrains the time of first light to be MJD $60410.56^{+0.07}_{-0.12}$. Both methods yield consistent explosion epochs to that derived above from the phases of last non-detection and first detection. All phases reported in this paper are with respect to this adopted time of first light ($\delta t$). In this paper, we use a redshift-independent host-galaxy distance of $7.24 \pm 0.20$~Mpc \citep{Saha06} and adopt a redshift of $z = 0.002215$ based on \ion{Na}{i}\ D absorption in high-resolution spectra of \sn{} obtained with the NEID Spectrograph on the WIYN Telescope (private communication). 

Given detection and classification during its infancy, \sn{} represents an incredible opportunity to study the SBO phase of a CSM-interacting SN~II in unprecedented detail. In \S\ref{sec:obs} we describe UV, optical, and NIR observations of \sn{}. In \S\ref{sec:analysis} we present analysis, comparisons and modeling of \sn{}'s optical photometric and spectroscopic properties. Finally, in \S\ref{sec:discussion} we discuss the progenitor environment and mass-loss history prior to \sn{}. Conclusions are drawn in \S\ref{sec:conclusion}. All uncertainties are quoted at the 68\% confidence level (c.l.) unless otherwise stated.


\section{Observations} \label{sec:obs}


\subsection{Photometric Observations}\label{SubSec:Phot}

The Ultraviolet Optical Telescope (UVOT; \citealt{Roming05}) onboard the Neil Gehrels \emph{Swift} Observatory \citep{Gehrels04} started observing \sn{} on 11 April 2024 ($\delta t = 0.79$~days). We performed aperture photometry with a 5$\arcsec$ region radius with \texttt{uvotsource} within HEAsoft v6.33 \citep{HEAsoft}\footnote{We used the most recent calibration database (CALDB) version.}, following the standard guidelines from \cite{Brown14}\footnote{\url{https://github.com/gterreran/Swift_host_subtraction}}. In order to remove contamination from the host galaxy, we employed pre-explosion images to subtract the measured count rate at the location of the SN from the count rates in the SN images and corrected for point-spread-function (PSF) losses following the prescriptions of \cite{Brown14}. We note that not all early-time UVOT observations are included in this analysis given the degree of saturation. 

We obtained $ugri$ imaging of \sn{} with the Las Cumbres Observatory (LCO) 1\,m telescopes from 12 April to 26 April 2024 (Program ANU2024A-004). After downloading the {\tt BANZAI}-reduced images from the LCO data archive \citep{mccully18}, we used {\tt photpipe} \citep{Rest+05} to perform {\tt DoPhot} PSF photometry \citep{Schechter+93}. All photometry was calibrated using PS1 stellar catalogs described above with additional transformations to SDSS $u$-band derived from \citet{finkbeiner16}. For additional details on our reductions, see \citet{kilpatrick18}. We also obtained photometry using a 0.7 meter Thai Robotic Telescope at Cerro Tololo Inter-American Observatory in the $ugriz$ bands. Images are bias subtracted and field flattened. Absolute photometry is obtained using stars in the 10$'\times$10$'$ field of view. We also observed \sn{} with the Lulin 1\,m telescope in $griz$ bands. Standard calibrations for bias and flat-fielding were performed on the images using {\tt IRAF}, and we reduced the calibrated frames in {\tt photpipe} using the same methods described above for the LCO images. 

\begin{figure*}
\centering
\subfigure{\includegraphics[width=0.50\textwidth]{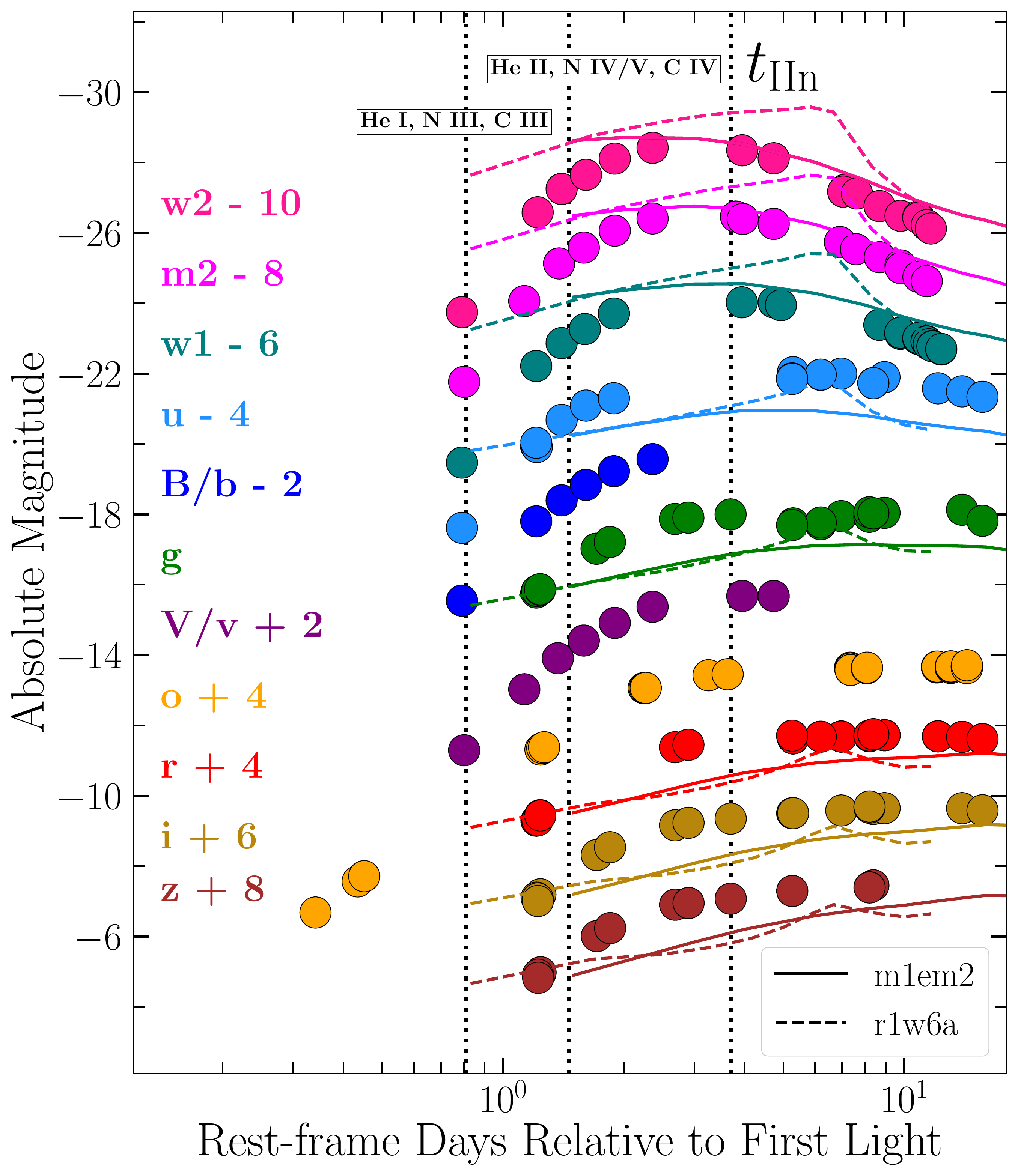}}
\subfigure{\includegraphics[width=0.49\textwidth]{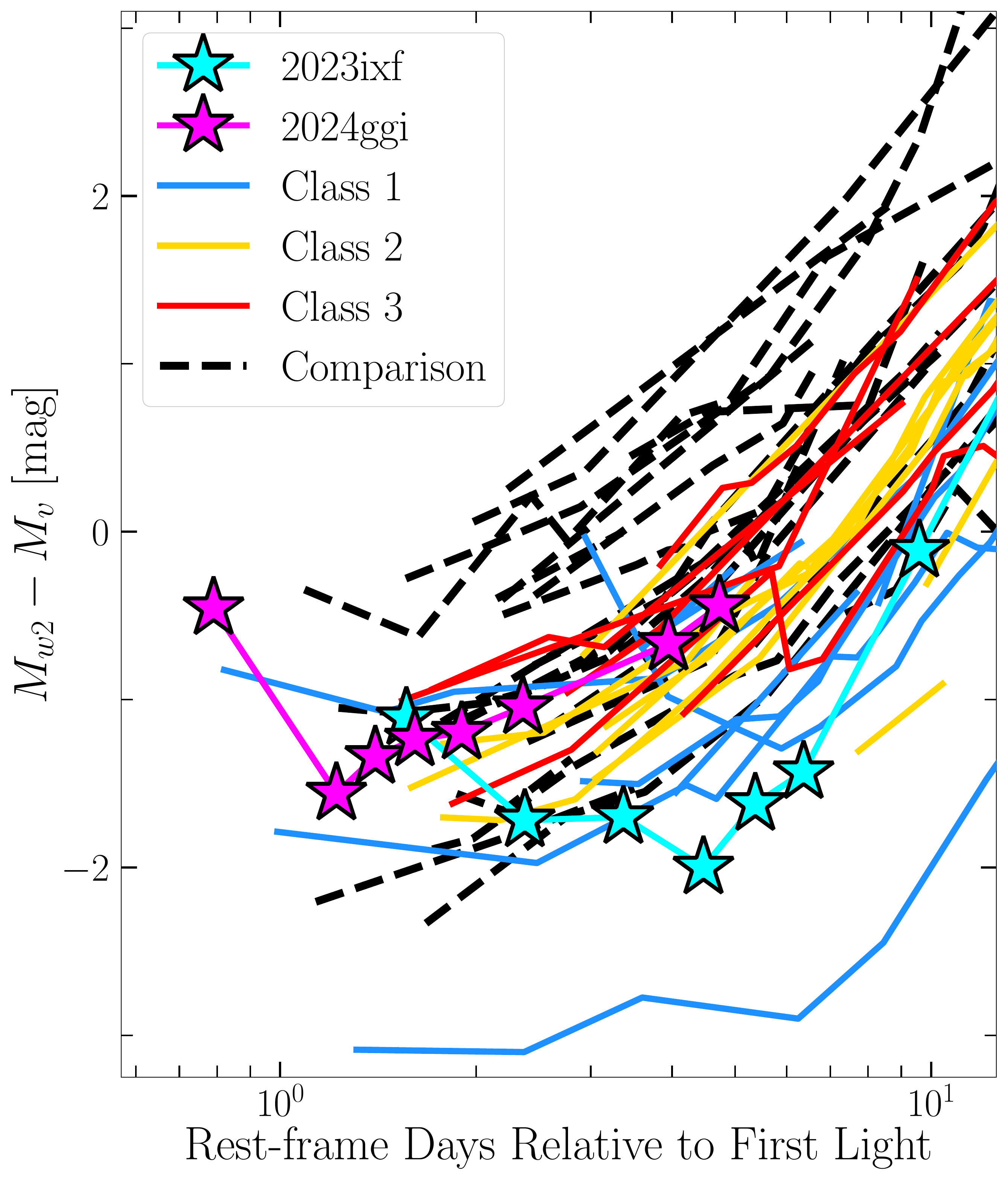}}
\caption{ {\it Left:} Multi-color light curve of \sn{} (circles) with respect to time since first light (MJD $60410.80 \pm 0.34$) from {\it Swift}, ATLAS, LCO, TRT, REM, T80s, and Lulin telescopes. Observed photometry is presented in the AB magnitude system and has been corrected for host galaxy and MW extinction. {\tt CMFGEN} m1em2 and r1w6a model light curves are shown as solid and dashed lines, respectively. {\it Right:} Early-time, reddening-corrected $W2-V$ color plot for \sn{} (magenta stars) and SN~2023ixf (cyan stars) with respect to gold/silver-sample objects (red, yellow, blue lines) and comparison sample objects (black dashed lines) \citep{wjg24}. \sn{} shows a blueward shift in color within the first $\sim$day since first light that is consistent with a rise in temperature and ionization (e.g., Fig. \ref{fig:spec_series}). 
\label{fig:LC_colors} }
\end{figure*}

We also observed \sn{} in $grizH$-bands with the Rapid Eye Mount (REM; \citealt{Antonelli2003}) telescope located in La Silla, Chile. REM is equipped with two cameras, which can observe simultaneously the same field of view (10'$\times$10') in the optical and near-IR. Single images have been initially corrected for dark and flat frames observed each night. Then, images obtained with the same setup have been stacked and finally corrected for cosmic rays, all using a dedicated pipeline written in Python. Magnitudes were measured with aperture photometry, with a variable aperture size according to the seeing of the night, and calibrated against selected field stars from the Skymapper DR4 \citep{Onken2024} that also have 2MASS $JHK$-band photometry. We also obtained $ugriz$ imaging of \sn{} with the 0.8\,m T80S telescope at Cerro Tololo Inter-american Observatory, Chile.  These were processed using the S-PLUS Transient Extension Program pipeline \citep[][]{Santos24}, including non-linearity to recover detections of SN\,2024ggi close to the saturation level. Additionally, we include $o$-band photometry by ATLAS that was downloaded from the forced photometry server \citep{tonry18, Smith20, Shingles21}. The complete multi-color light curve of \sn{} is presented in Figure \ref{fig:LC_colors}.

The Milky Way (MW) $V$-band extinction and color excess along the SN line of sight is $A_{V} = 0.22$~mag and \textit{E(B-V)} = 0.07~mag \citep{schlegel98, schlafly11}, respectively, which we correct for using a standard \cite{fitzpatrick99} reddening law (\textit{$R_V$} = 3.1). In addition to the MW color excess, we estimate the contribution of galaxy extinction in the local SN environment. Using a high-resolution Kast spectrum of \sn{} at $\delta t = 5.5$~days, we calculate \ion{Na}{i} D2 and D1 equivalent widths (EWs) of 0.18 and 0.13~\AA, respectively; these values are confirmed in a follow-up high resolution Gemini spectrum at $\delta t = 9.2$~days. We use $A_V^{\rm host} = (0.78\pm0.15)~{\rm mag} \times ({\rm EW_{NaID}}$/\AA) from \cite{Stritzinger18} to convert these EWs to an intrinsic host-galaxy $E(B-V)$ and find a host galaxy extinction of $E(B-V)_{\textrm{host}} = 0.084 \pm 0.018$~mag, also corrected for using the \cite{fitzpatrick99} reddening law.

\subsection{Spectroscopic Observations}\label{SubSec:Spec}

\begin{figure*}[t]
\centering
\includegraphics[width=0.99\textwidth]{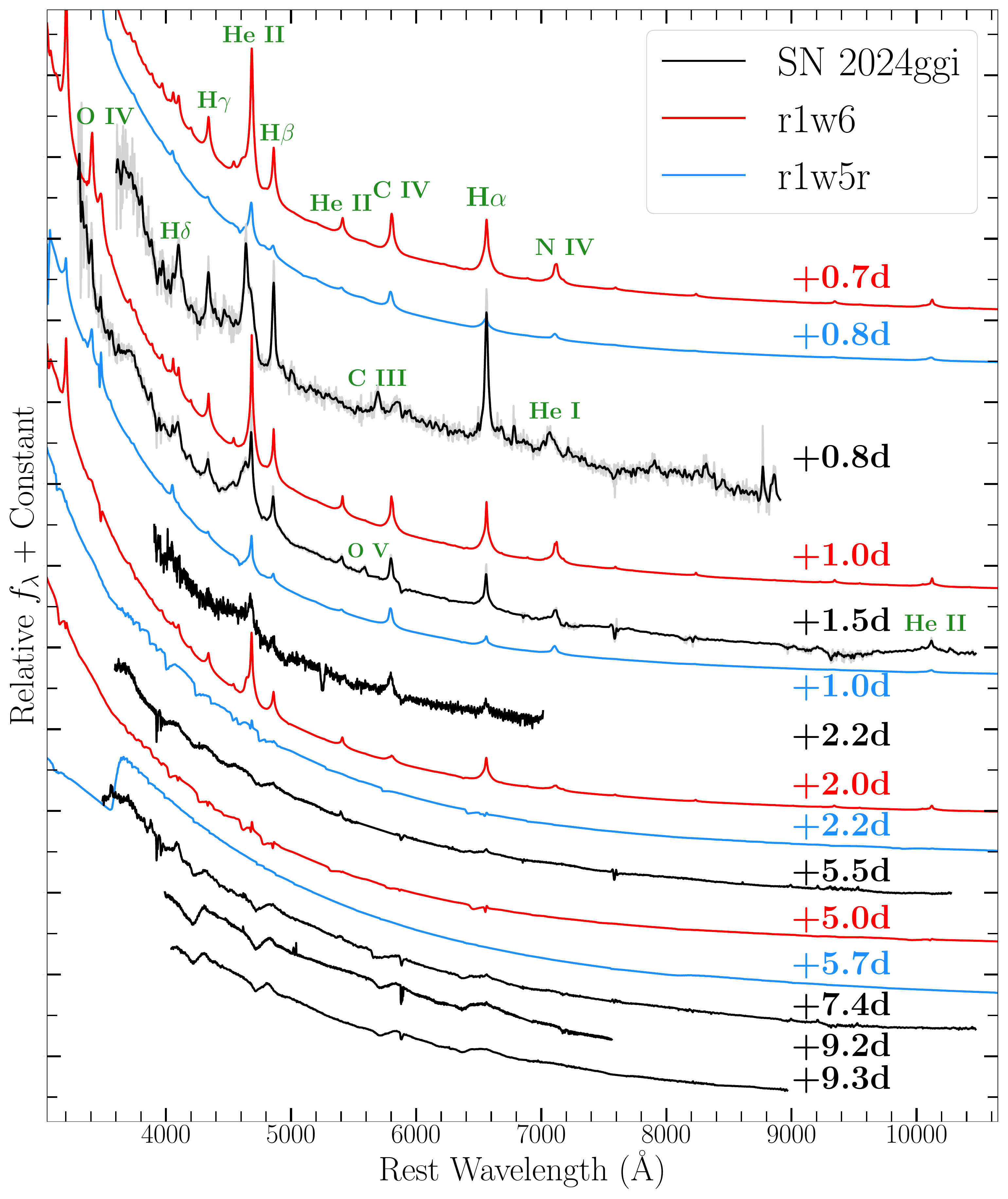}
\caption{Early-time optical spectral series of \sn{} (black) together with r1w6 (red) and r1w5r (blue) models from \cite{dessart17}. The r1w6 and r1w5r models are characterized by wind mass-loss rates of $\dot{M} = (0.5-1) \times 10^{-2}~\Msun$~yr$^{-1}$ that extends to a CSM radius of $R_{\rm CSM} = (2-5)\times 10^{14}$~cm. Model spectra have been smoothed with a Gaussian kernel to match the spectral resolution of the data. Line identifications shown in green. The appearance of \ion{He}{ii}, \ion{C}{iv}, \ion{N}{iv/v} and \ion{O}{v} after the $\delta t = +0.8$~day spectrum indicates a rise in ionization and temperature in \sn{} as the breakout pulse and the subsequent continuous release of radiation from the shock diffuses through the CSM. \label{fig:spec_series} }
\end{figure*}

\begin{figure*}
\centering
\subfigure{\includegraphics[width=0.49\textwidth]{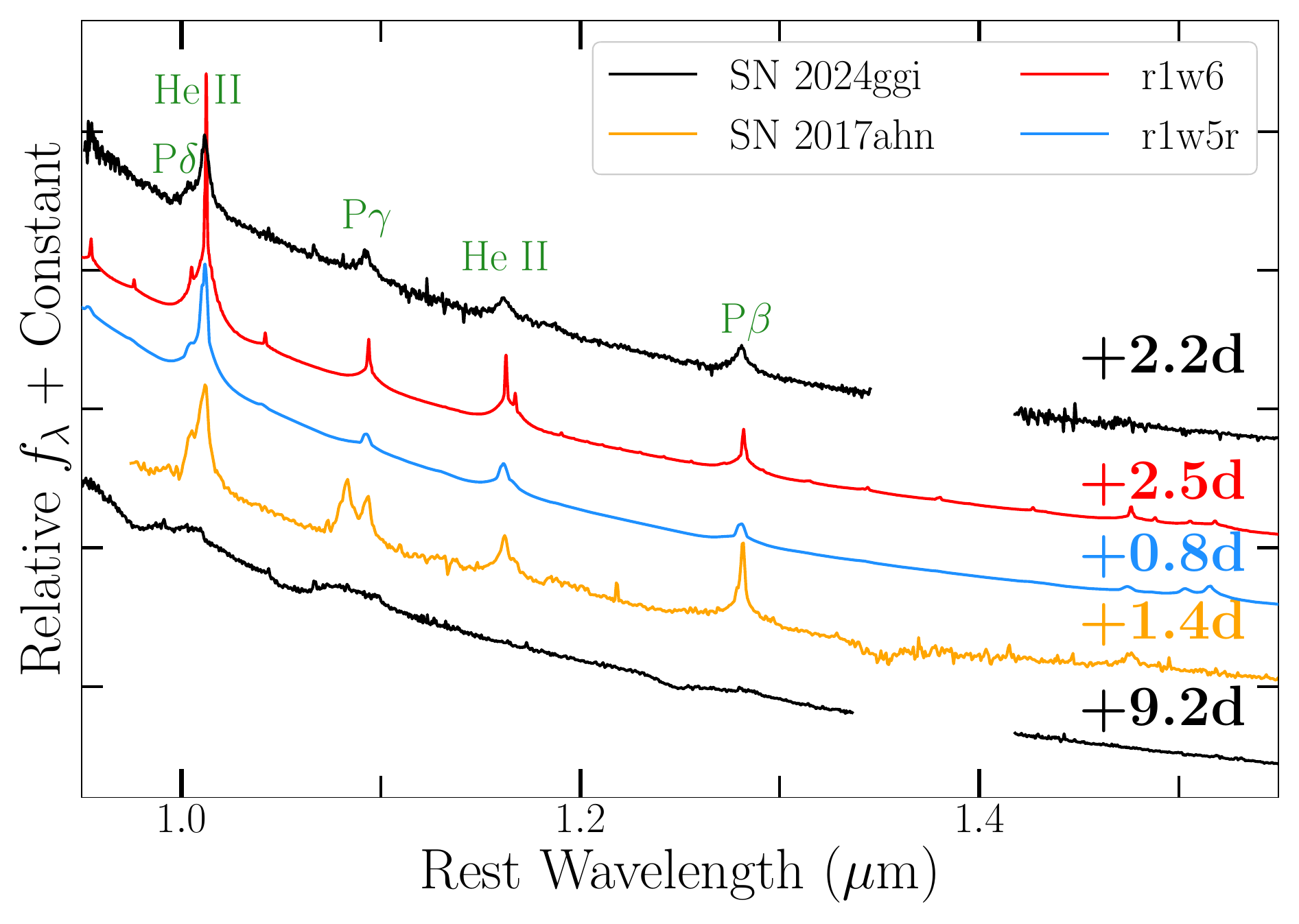}}
\subfigure{\includegraphics[width=0.49\textwidth]{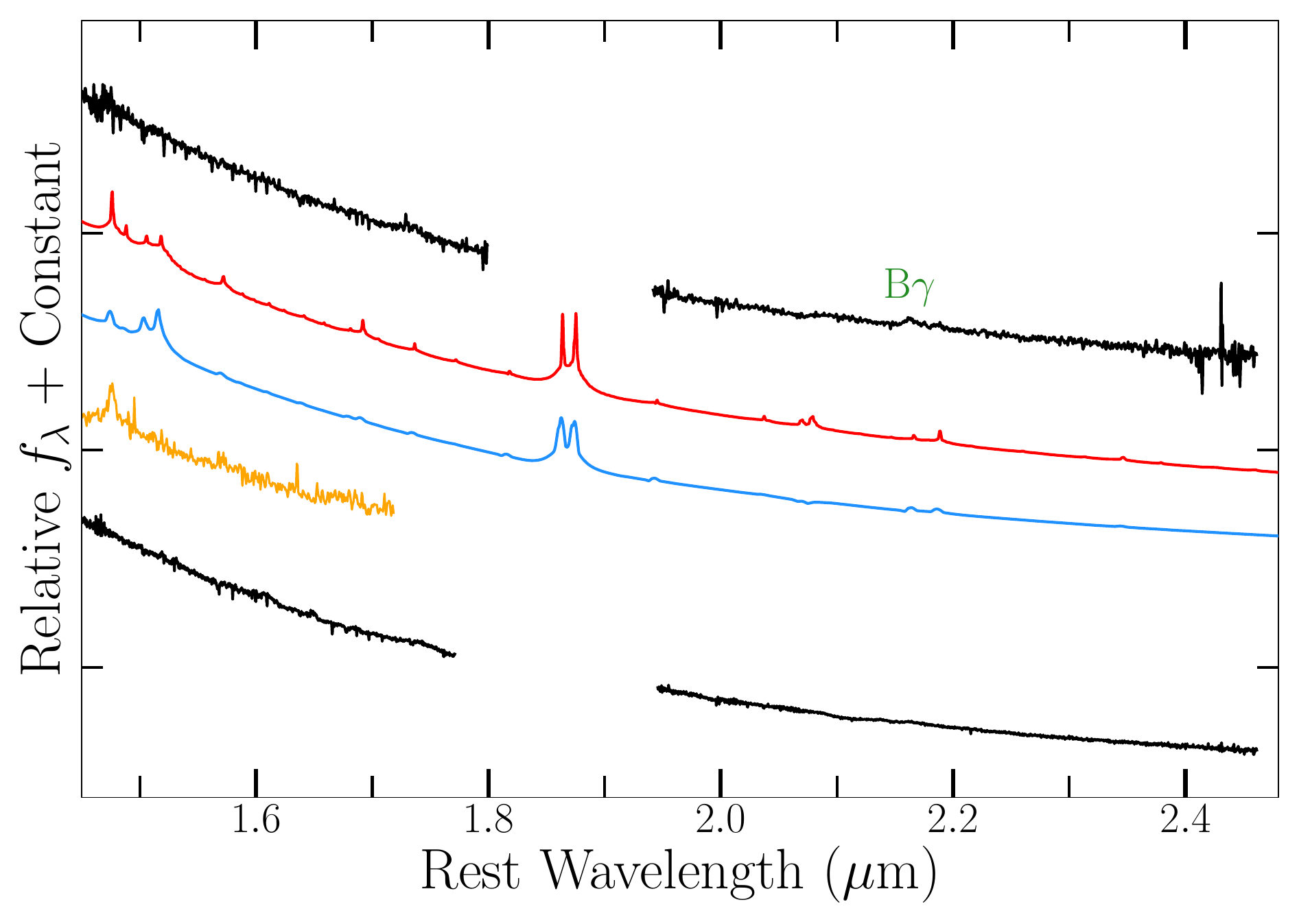}}
\caption{ {\it Left/Right:} NIR spectra of \sn{} (black) and SN~2017ahn (orange) compared to r1w5r (blue) and r1w6 (red) models from \cite{dessart17}. Line identifications shown in green. 
\label{fig:nir_spec} }
\end{figure*}

\sn{} was observed with Shane/Kast \citep{KAST} and the Goodman spectrograph \citep{clemens04} at the Southern Astrophysical Research (SOAR) telescope between $\delta t = 1.5 - 9.3$~days. For all these spectroscopic observations, standard CCD processing and spectrum extraction were accomplished with \textsc{IRAF}\footnote{\url{https://github.com/msiebert1/UCSC\_spectral\_pipeline}}. The data were extracted using the optimal algorithm of \citet{1986PASP...98..609H}.  Low-order polynomial fits to calibration-lamp spectra were used to establish the wavelength scale and small adjustments derived from night-sky lines in the object frames were applied. \sn{} spectra were also obtained Gemini Multi-Object Spectrograph (GMOS) at Gemini South Observatory at $\delta t = 9.2$~days and reduced with the Data Reduction for Astronomy from Gemini Observatory North and South (DRAGONS) pipeline \citep{Labrie23}. Spectra were also obtained with the TripleSpec4.1 NIR Imaging Spectrograph (TSpec) at the SOAR telescope, and reduced using a modified version of {\tt Spextool} \citep{CVR2004}. Telluric corrections were applied using {\tt xtellcor} presented in \citet{VCR2003}. Additional modifications to calibrations are described in \citet{Kirkpatrick2011}.

In Figure \ref{fig:spec_series} we present the complete series of optical spectroscopic observations of \sn{} from $\delta t = 0.8-9.3$~days. In this plot, we also show the classification spectrum of \sn{} at $+0.8$~days from the Lijiang 2.4m telescope \citep{Zhai24}, which we only use for narrow line identification. The complete optical/NIR spectral sequence is shown in Figures \ref{fig:spec_series} \& \ref{fig:nir_spec} and the log of spectroscopic observations is presented in Appendix Table \ref{tab:spec_table}.


\section{Analysis}\label{sec:analysis}

\subsection{Photometric Properties}\label{subsec:phot_properties}

We present the complete early-time, multi-band light curve of \sn{} in Figure \ref{fig:LC_colors}. Given the estimated time of first light, \sn{} was first detected by ATLAS at $\delta t = 0.3$~days with absolute magnitude of $M_o = -10.7$~mag and then quickly increased in luminosity to  $M_o = -15.3$~mag by $\delta t = 1.2$~days. We fit high-order polynomials to $w2,m2,w1,u,g,r,i$-band light curves in order to estimate the peak luminosity and rise-time of \sn{}. All measurements are reported in Table \ref{tbl:params}, with the uncertainty in peak magnitude being the $1\sigma$ error from the fit and the uncertainty in the peak phase being found from adding the uncertainties in both the time of peak magnitude and the time of first light in quadrature. We find that \sn{} has UV and optical peak absolute magnitudes of $M_{w2} = -18.7 \pm 0.07$~mag and $M_{g} = -18.1 \pm 0.06$~mag, respectively. Using the adopted time of first light, we estimate UV and optical rise-times of $t_{w2} = 3.0 \pm 0.3$~days and $t_{g} = 6.5 \pm 0.9$~days.   

\begin{figure*}
\centering
\subfigure{\includegraphics[width=0.49\textwidth]{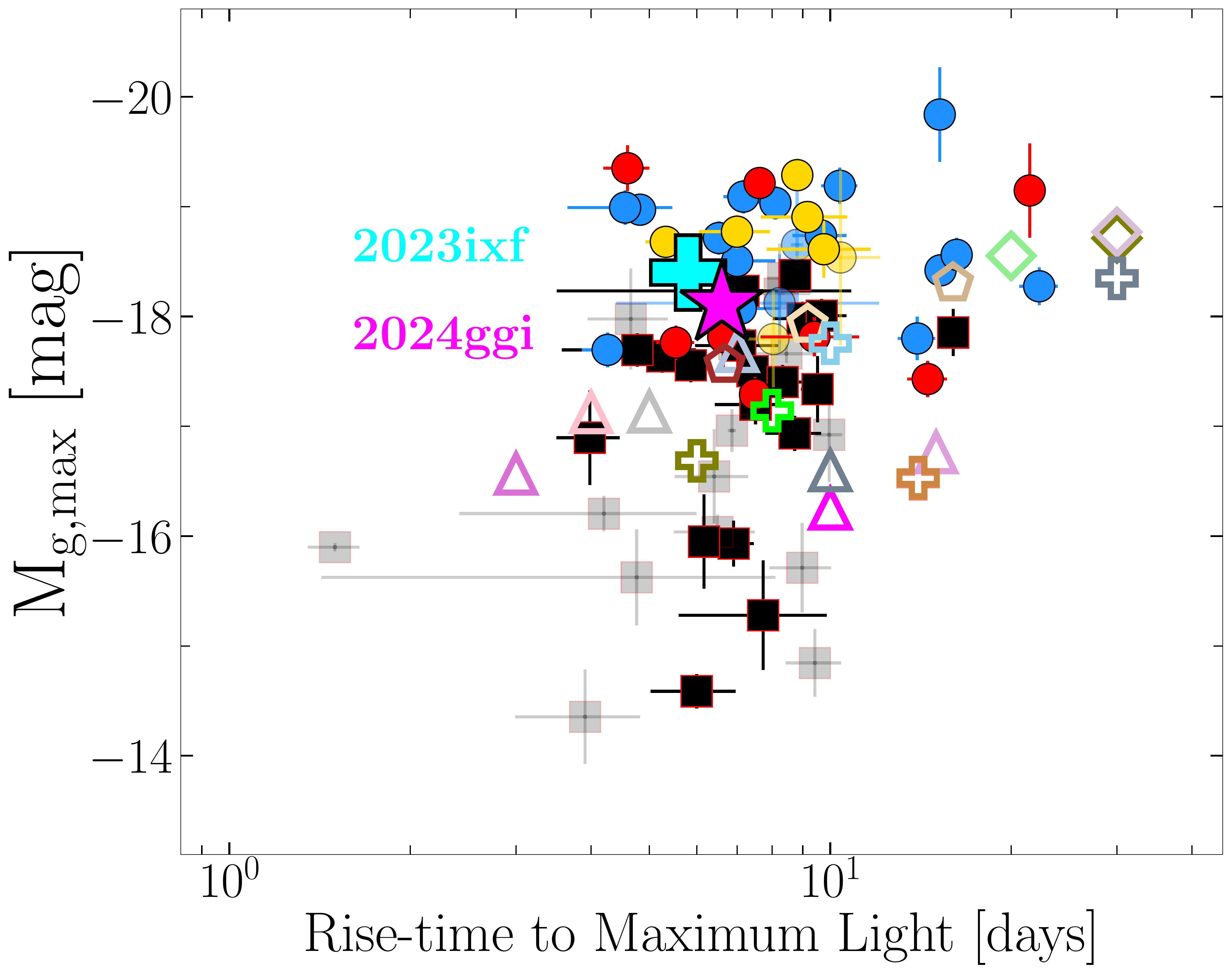}}
\subfigure{\includegraphics[width=0.49\textwidth]{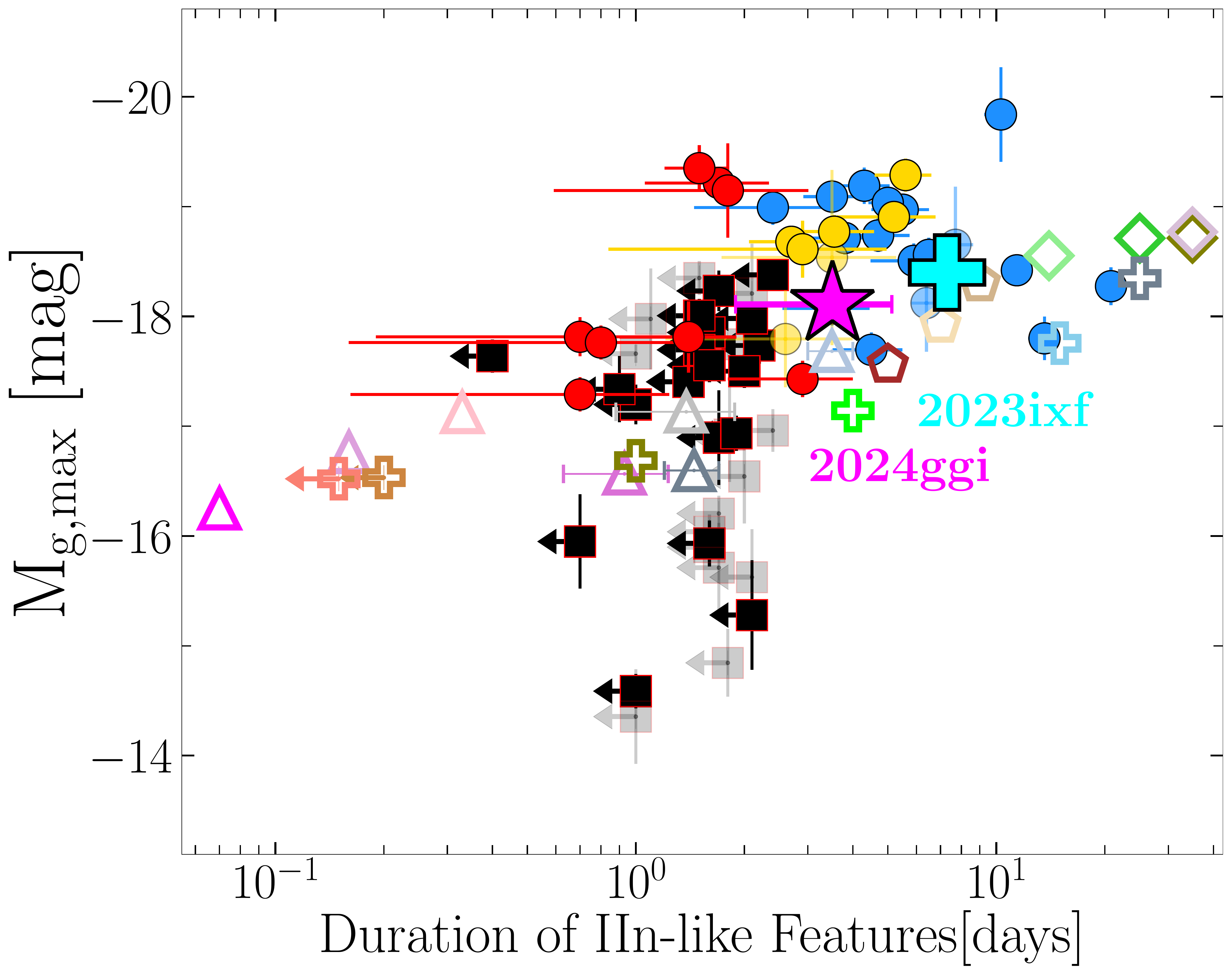}}\\
\subfigure{\includegraphics[width=0.99\textwidth]{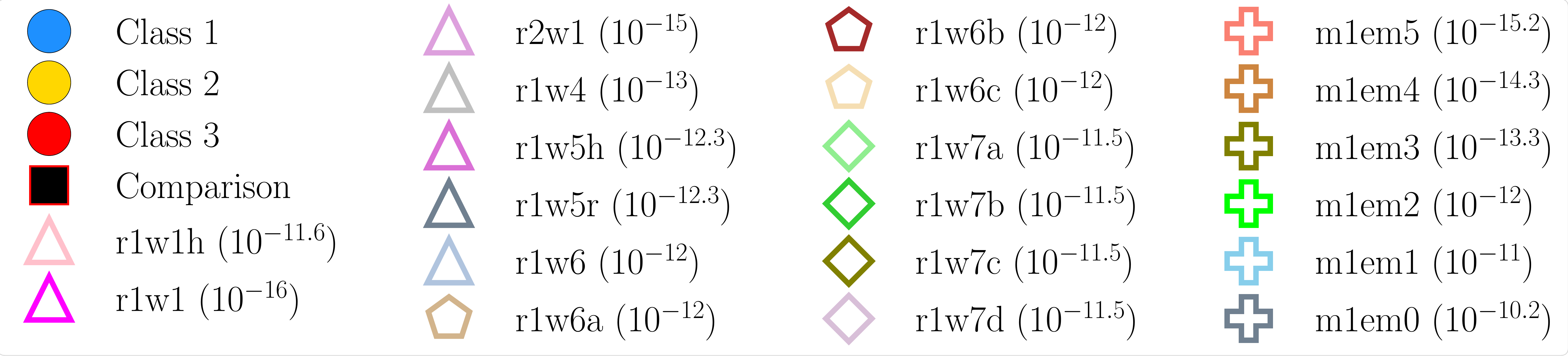}}
\caption{ {\it Left:} Peak \textit{g}-band absolute magnitude versus rise time (Left) and duration of IIn-like features (Right). Gold/silver samples shown as blue/yellow/red circles and the comparison sample is shown as black squares. Solid colored points represent the subsample of objects at $D>40$~Mpc. Parameters from the \cmfgen\ model grid (\S\ref{subsec:modeling}) are plotted as colored stars, polygons, diamonds and plus signs with the CSM densities at $10^{14}$~cm (in g~cm$^{-3}$) for each model displayed in parentheses. SNe~2024ggi and 2023ixf are shown as a magenta star and a cyan cross, respectively. 
\label{fig:max_lines} }
\end{figure*}

In Figure \ref{fig:max_lines}, we compare the observed peak absolute magnitudes of \sn{} to a sample of 74 SNe~II from \cite{wjg24}. This sample includes 39 SNe~II with detected IIn-like features in their early-time spectra: ``gold-sample'' objects have spectra at $\delta t < 2$~days and ``silver-sample'' objects only have spectra obtained at $\delta t > 2$~days. As discussed in \cite{wjg24} and delineated by color in Figure \ref{fig:max_lines}, the gold/silver-sample objects are classifed in three main groups: ``Class 1'' (blue) show emission lines of \ion{N}{iii}, \ion{He}{ii}, and \ion{C}{iv} (e.g., SNe~1998S, 2017ahn, 2018zd, 2020pni, 2020tlf, 2023ixf), ``Class 2'' (yellow) have no \ion{N}{iii} emission but do show \ion{He}{ii} and \ion{C}{iv} (e.g., SNe~2014G, 2022jox), and ``Class 3'' (red) only show weaker, narrow \ion{He}{ii} emission superimposed with a blueshifted, Doppler-broadened \ion{He}{ii} (e.g., SN~2013fs, 2020xua). However, this classification scheme is epoch dependent because emission lines of \ion{O}{v/vi} and \ion{N}{iv/v} are also present in some objects such as SN~2013fs at $t < 1$~day owing to a more compact CSM than other CSM-interacting SNe~II \citep{yaron17, dessart17}. Additionally, we present the comparison sample of 35 SNe~II with spectra obtained at $\delta t < 2$~days but no detected IIn-like features.

As shown in Figure \ref{fig:max_lines}, \sn{} is more luminous than most in the comparison sample SNe~II without IIn-like features in all UV/optical filters. Furthermore, \sn{} has a longer rise-time in UV filters than comparison sample objects but comparable rise-times in optical filters. However, \sn{} shows consistent UV/optical luminosities and rise-times to gold/silver-sample objects such as iPTF11iqb \citep{smith15} and SN~2014G \citep{Terreran16}. Additionally, \sn{} has a comparable optical luminosity to SN~2023ixf but is less luminous in the UV by $\sim$0.7~mags. In the right panel of Figure \ref{fig:LC_colors}, we also compare the early-time ($\delta t < 10$~days) $w2-v$ colors of \sn{} to gold, silver and comparison objects. Interestingly, \sn{} shows a dramatic red-to-blue $w2-v$ color evolution of $-0.46$~mag to $-1.56$~mag between $\delta t = 0.8 - 1.2$~days, followed by consistently blue colors as it evolves redward in its first week. This unusual color evolution is also observed in other SNe~II with IIn-like features such as SN~2023ixf \citep{Hosseinzadeh23, Hiramatsu23, 
Li24, Zimmerman24}, which was proposed as evidence for SBO in an extended dense CSM. This phenomenon is a product of the breakout pulse and the subsequent continuous release of radiation from the shock diffusing through the CSM \citep{dessart17} and also corresponds to the phase during which the photosphere moves outward, initially at $R_{\star}$ and then out to the location of the electron scattering photosphere ($R_{\tau = 1}$; \citealt{Dessart23}). We also fit a blackbody model to the \sn{} UV/optical SED and find blackbody temperature[radii] during this red-to-blue color evolution of $\sim$20~kK[$5.3\times10^{13}$~cm] at $\delta t = 0.8$~days and $\sim$23~kK[$1.1\times10^{14}$~cm] at $\delta t = 1.2$~days. Furthermore, we find that the blackbody temperature of \sn{} peaks at $\sim$34~kK on $\delta t = 1.4$~days. This is a similar peak blackbody temperature to the 34.3~kK found for SN~2023ixf at $\delta t = 3.51$~days \citep{Zimmerman24}. However, we note that these blackbody temperatures correspond to the temperature at the thermalization depth and that the blackbody radius is not equivalent to the location of the photosphere (i.e., $R_{\rm BB} << R_{\rm phot}$).

\subsection{Spectroscopic Properties}\label{subsec:spec_analysis}

\begin{figure*}
\centering
\subfigure{\includegraphics[width=0.33\textwidth]{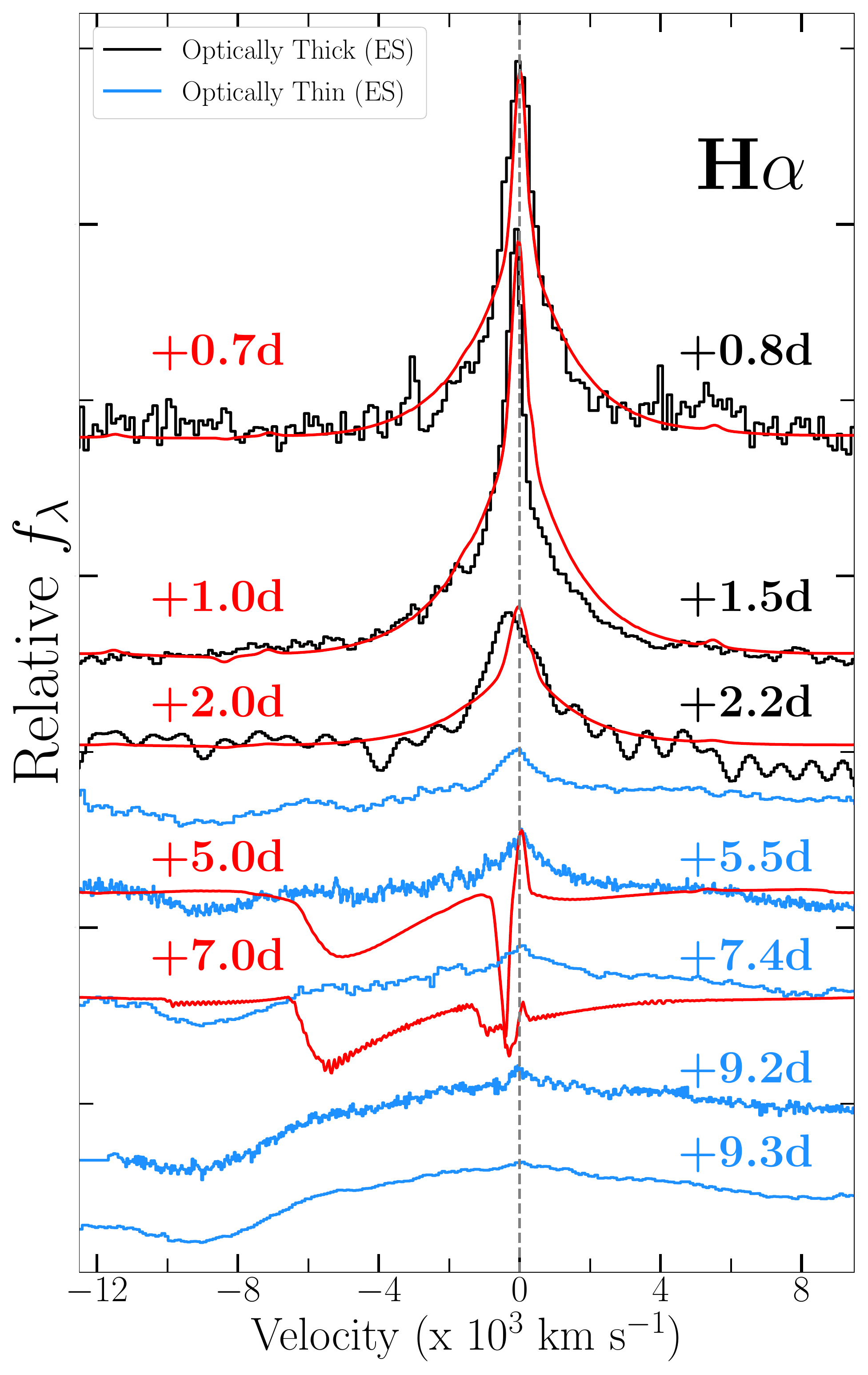}}
\subfigure{\includegraphics[width=0.33\textwidth]{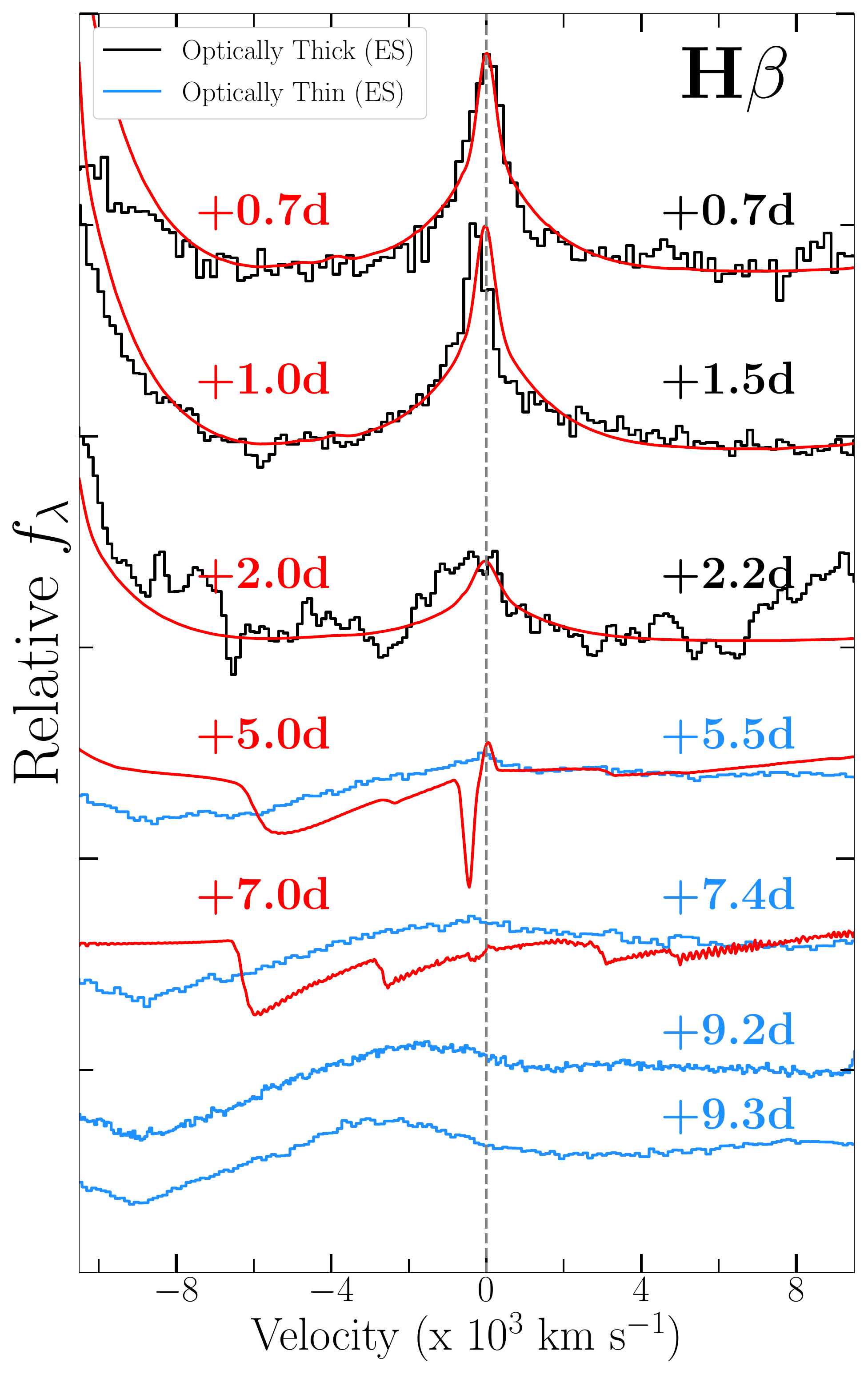}}
\subfigure{\includegraphics[width=0.33\textwidth]{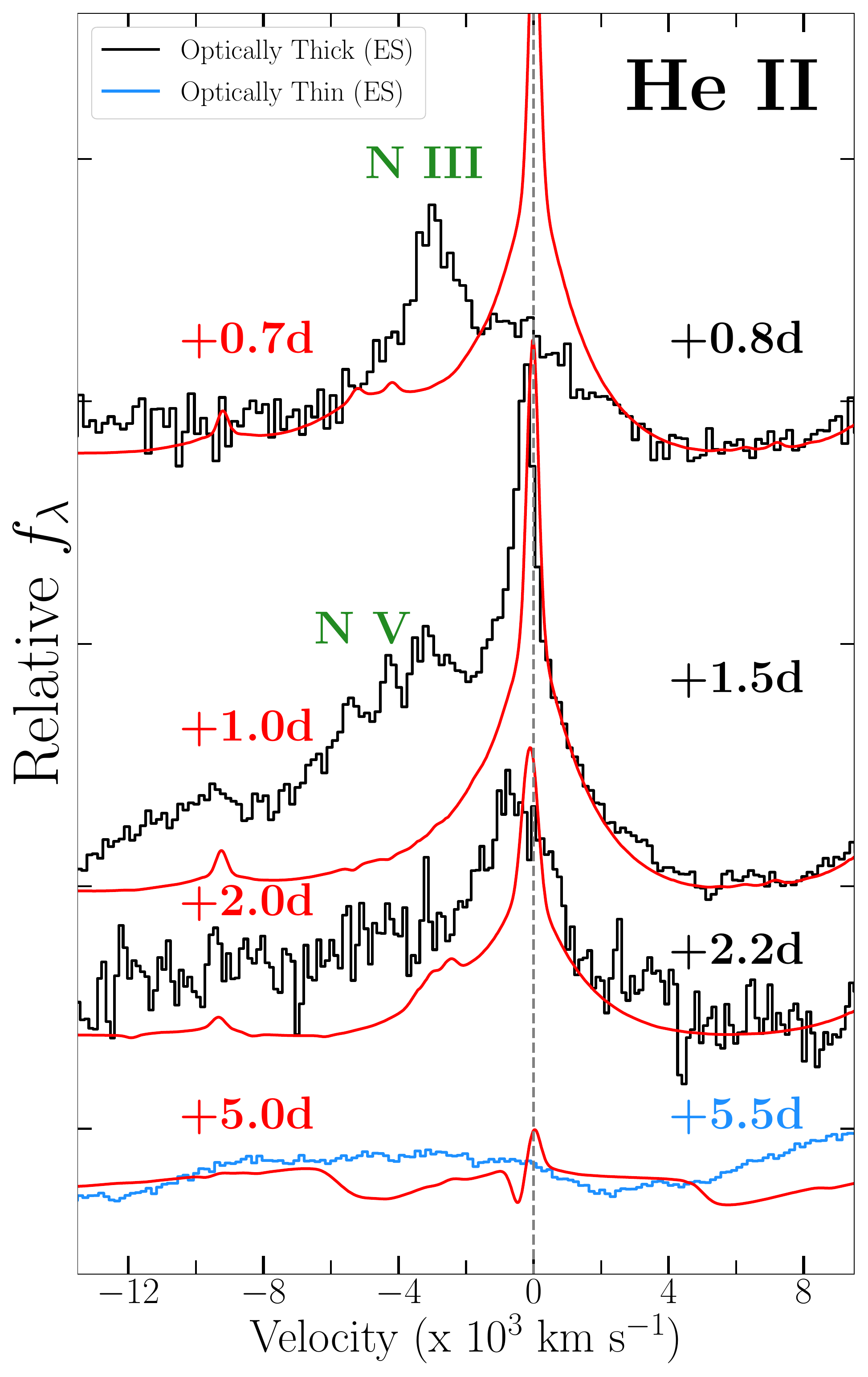}}
\caption{{\it Left:} H$\alpha$ velocity evolution in \sn{} (black) from $\delta t = 0.8 - 9.3$~days with respect to r1w6 model spectra (red), which has been scaled to the emission line peaks of \sn{} and smoothed with a Gaussian filter to better compare with the data. Early-time spectral profiles are shaped by electron-scattering in the dense CSM. The transition shown from black to blue lines ($t_{\rm IIn} = 3.7 \pm 1.8$~days) marks the emergence of broad absorption features derived from the fastest moving SN ejecta. {\it Middle:} H$\beta$ velocity evolution, also showing that the electron-scattering line profiles subside by $\sim 5.5$~days. {\it Right:} \ion{He}{ii}~$\lambda$4686 velocity evolution reveals that the electron-scattering profile fades by $\sim 5.5$~days, suggesting a significant decrease in CSM density.   
\label{fig:vels} }
\end{figure*}  

We present our sample of optical observations for \sn{} spanning from $\delta t = +0.8$ to $+9.3$~days in Figure \ref{fig:spec_series}. In the earliest spectrum at $\delta t = +0.8$~days, \sn{} shows IIn-like features of \ion{H}{i} ($\chi = 13.6$~eV), \ion{He}{i} ($\chi = 24.6$~eV), \ion{N}{iii} ($\chi = 47.4$~eV) and \ion{C}{iii} ($\chi = 47.9$~eV). However, by the next spectral observation at $\delta t = +1.5$~days, \sn{} shows prominent emission lines of \ion{He}{ii} ($\chi = 54.5$~eV), \ion{N}{iii/iv/v} ($\chi = 47.4/77.5/97.9$~eV), \ion{C}{iv} ($\chi = 64.5$~eV) and  \ion{O}{v} ($\chi = 113.9$~eV), indicating a dramatic rise in ionization and temperature within $\sim$14~hours that is temporally consistent with the blueward evolution in $w2-v$ colors (e.g., Fig. \ref{fig:LC_colors}). There may also be a detection of \ion{O}{iv} $\lambda3410$ but the lower S/N in that spectral region makes the line identification uncertain. Interestingly, the detection of \ion{O}{iv/v} has only been confirmed in one other SN~II, 2013fs, and in that object the timescale of this emission was $<0.5$~days \citep{yaron17}. Notably, the shift in ionization seen in \sn{} was also observed in SN~2023ixf but at later phases of $+1.1 - 2.4$~days \citep{wjg23}. A two-component Lorentzian model fit to the H$\alpha$ profile in the +1.5~d Kast spectrum shows a narrow component full width at half maximum (FWHM) velocity of $<$270~$\kms$ that traces the unshocked CSM combined with a broad symmetric component with velocity of $\sim$1320~$\kms$ that is caused by scattering of recombination line photons by free, thermal electrons in the ionized CSM \citep{Chugai01, Dessart09, Huang18}. However, the narrow component is likely affected by radiative acceleration of the CSM that causes the width of the narrow component to be larger than the true velocity of the progenitor wind \citep{D15_2n, dessart17,Tsuna23}. 

In addition to optical spectroscopy, we present NIR spectra of \sn{} in Figure \ref{fig:nir_spec} that extend from $0.9-2.4~\mu$m. The first spectrum at $\delta t = 2.2$~days shows IIn-like features of \ion{He}{ii} as well as Hydrogen Paschen and Brackett transitions of \ion{P}{$\delta$}, \ion{P}{$\gamma$}, \ion{P}{$\beta$} and \ion{B}{$\gamma$}. We then compare this spectrum to SN~II 2017ahn \citep{Tartaglia21}, one of the only SN with IIn-like features to have a NIR spectrum during the dense CSM-interaction phase. Overall, the NIR spectrum of \sn{} at $\delta t = 2.2$~days shows similar narrow emission lines to SN~2017ahn with the exception being that SN~2017ahn shows prominent \ion{He}{i} emission while \sn{} only shows \ion{He}{ii}. 

As shown in Figure \ref{fig:vels}, the IIn-like features in \sn{} are already fading by the +2.2~day epoch, with possible blue-shifted, Doppler-broadened \ion{He}{ii} emission as the fastest moving ejecta and/or the dense shell starts to become visible given a decrease in optical depth. Then, by $\delta t = 5.5$~days, the IIn-like features have vanished and \sn{} shows a broad, blueshifted absorption profile in all Balmer transitions as well as a blue-shifted \ion{He}{ii} profile extending out to $\sim (10-12)\times 10^4~\kms$ (i.e., ``ledge feature''; \citealt{Dessart16,Hosseinzadeh22, Pearson23, Chugai23, Shrestha24}). Based on this evolution, we estimate the duration of the IIn-like line profiles to be $t_{\rm IIn} = 3.8 \pm 1.6$~days, which marks the transition point at which the optical depth to electron-scattering has dropped enough to see the emerging fast-moving SN ejecta. At $\delta t > 5$~days, we observe broad absorption profiles in the H$\alpha$ and H$\beta$ transitions that extend to $\sim 12000~\kms$, which provides a decent estimate of the velocities of the fastest moving H-rich material at the shock front. Using the estimated $t_{\rm IIn}$ and a shock velocity of $v_{\rm sh} = 12000~\kms$, the transition to lower density CSM likely occurs at a radius of $r = R_{\star} + v_{\rm sh} \times t_{\rm IIn} = (4.3 \pm 1.7) \times 10^{14}$~cm (for $R_{\star} = 500~\Rsun$) and at a CSM density of $\rho = (\kappa_{\rm T} \times t_{\rm IIn} \times v_{\rm sh})^{-1} = (7.4 \pm 3.1) \times 10^{-15}$~g~cm$^{-3}$, for $\tau = 1$ and Thompson opacity of $\kappa_{\rm T} = 0.34$~cm$^2$~g$^{-1}$. This likely indicates a more confined CSM and/or lower mass-loss rate for \sn{} than Class 1 gold/silver sample SNe~II that have $t_{\rm IIn} > 5$~days e.g., SNe~2017ahn, 2018zd, 2020pni, 2020tlf, 2023ixf. However, as shown in Figure \ref{fig:max_lines}, the duration of IIn-like features in \sn{} is similar to Class 2/3 objects, which reveals that if earlier time spectra (e.g., $\delta t < 1$~day) had been obtained for these objects, they may have shown \ion{N}{iii} emission like Class 1 objects and \sn{} at $\delta t < 1$~day. 

\subsection{Model Matching}\label{subsec:modeling}

\begin{figure}[t!]
\centering
\includegraphics[width=0.47\textwidth]{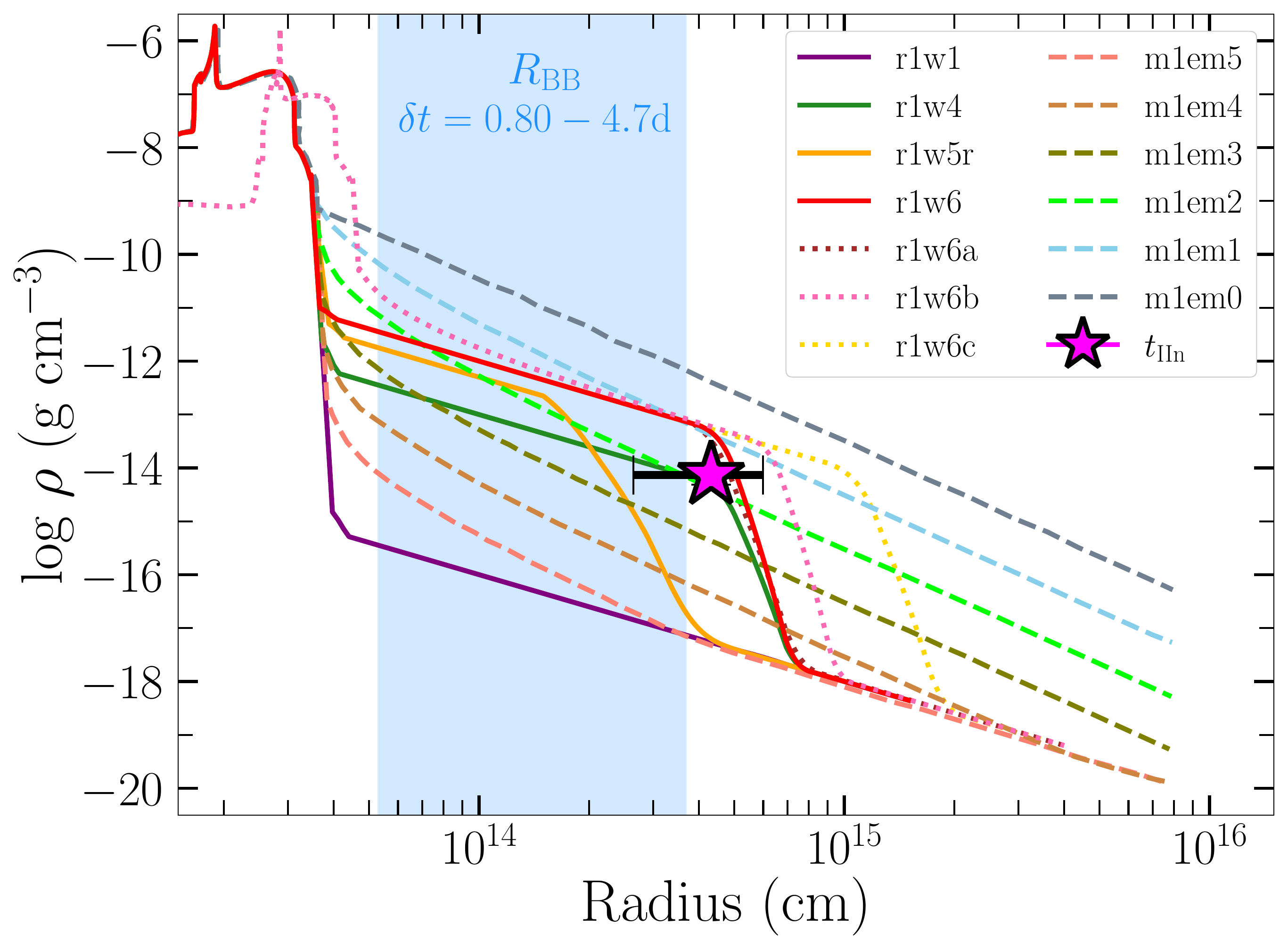}
\caption{CSM densities and radii for a subset of the \cmfgen\ model grid used in \cite{wjg24}. Best-matched models to \sn{} (\S\ref{subsec:modeling}) are r1w6 (solid red line), m1em2 (dashed limegreen line), r1w4 (solid dark green line) and r1w5r (solid orange line). Blackbody radii between $\delta t = 0.8 - 4.7$~days shown in light blue shaded region. Transition point where the CSM goes from optically thick to thin to electron scattering shown as magenta star (\S\ref{subsec:spec_analysis}). \label{fig:density}}
\end{figure}

In order to quantify the CSM properties of \sn{}, we compared the spectral and photometric properties of \sn{} to a model grid of radiation hydrodynamics and non-LTE radiative transfer simulations covering a wide range of progenitor mass-loss rates ($\dot{M} = 10^{-6} - 10^{0}~\Msun$~yr$^{-1}$; $v_w = 50~\kms$) and CSM radii ($R = 10^{14} - 10^{16}$~cm), all in spherical symmetry. Density profiles for a representative set of models are present in Figure \ref{fig:density}. Simulations of the SN ejecta-CSM interaction were performed with the multi-group radiation-hydrodynamics code \heracles\ \citep{gonzalez_heracles_07,vaytet_mg_11,D15_2n}, which consistently computes the radiation field and hydrodynamics. Then, at selected snapshots in time post-explosion, the hydrodynamical variables are imported into the non-LTE radiative-transfer code \cmfgen\ \citep{hillier12, D15_2n} for an accurate calculation of the radiative transfer, which includes a complete model atom, $\sim10^6$ frequency points, and treatment of continuum and line processes as well as non-coherent electron scattering. For each model, we adopt an explosion energy of $1.2\times 10^{51}$~erg, a 15$\Msun$ progenitor with a radius ranging from $R_{\star} \approx 500 - 700~\Rsun$, and a CSM composition set to the surface mixture of a RSG progenitor \citep{dessart17}. For the simulations presented in this work, the CSM extent is much greater than $R_{\star}$ ($\sim$500--1200~$\Rsun$ for a RSG mass range of $\sim$10--20~$\Msun$) and $R_{\star}$ has little impact during phases of ejecta-CSM interaction. The progenitor radius plays a more significant role on the light curve evolution during the plateau phase (e.g., see \citealt{d13_sn2p,wjg22}), i.e., once the interaction phase is over and the emission from the deeper ejecta layers dominate the SN luminosity. Specific methods for each simulation can be found in \cite{Dessart16, dessart17, wjg22, Dessart23}. 

We determine the best-matched model to \sn{} through direct spectral matching and peak/rise-time in all available UV/optical filters (see \citealt{wjg24} for specifics of the model matching procedure). As shown in Figure \ref{fig:spec_series}, the early-time evolution of \sn{} is most consistent with the r1w6 model, suggesting a mass-loss rate of $\dot M = 10^{-2}$~\mdot and a dense CSM (i.e., optically thick to electron scattering) that extends to $\sim 5\times 10^{14}$~cm. In Figure \ref{fig:vels} we show that the line profiles in \sn{} during the first few days of \sn{} are decently matched by the r1w6 model and that a lower mass-loss rate model (e.g., $\dot M = 5\times 10^{-3}$~\mdot; r1w5r) cannot match the duration of the IIn-like features. However, notable differences include the lack of \ion{N}{iii} and the strong \ion{He}{ii} at +0.8~days that is not present in \sn{} and inability of the r1w6 model to adequately reproduce the complete \ion{N}{iii}/\ion{He}{ii} complex i.e., the model profile is too narrow as was observed in SN~2023ixf \citep{wjg23}, which could be due to the fact that these simulations assume spherical symmetry and/or require higher kinetic energies. After $t_{\rm IIn}$, the r1w6 model spectra show a narrow P-Cygni profile combined with broad absorption that extends to $\sim7000~\kms$ in Balmer transitions. However, while this is qualitatively the behavior observed in \sn{}, there is no narrow P Cygni observed until the $+9.3$~day spectra and the absorption profiles in \sn{} extend to much larger velocities of $\sim 12000~\kms$. This mismatch is likely due to increased ejecta deceleration in the models that is not present in \sn{} and/or the CSM of \sn{} is asymmetric, allowing for some parts of the SN ejecta that only encounter low density CSM to retain higher velocities. Asymmetric CSM has been proposed for similar events such as SNe~1998S and 2023ixf \citep{Leonard00, Vasylyev23} based on spectropolarimetry. Similar to SN~2023ixf, CSM asymmetry may need to be invoked if the densities inferred from X-ray detections \citep{Margutti24,zhang24} deviate from those inferred from the early-time UV/optical light curve and spectra. Furthermore, it is expected that the dense shell formed from post-shock gas should break up, even for a spherical explosion and spherical CSM, which would lead to a much more complicated structure than is obtained when spherical symmetry is imposed. 

In addition to direct spectral comparison, we also use the peak absolute magnitudes and rise-times in all UV/optical filters to determine the best-matched models from the {\tt CMFGEN} grid. Using the residuals between the model predictions and observables, we find that \sn{} is best matched by the r1w4, m1em2, r1w6 and r1w6b models, which have a mass-loss range of $\dot M = (0.1-1)\times 10^{-2}$~\mdot and radii of dense CSM that ranges from $r = 4-8 \times 10^{14}$~cm. We note that the m1em2 model \citep{Dessart23} does not have a confined shell of dense CSM but rather has a continuous $r^{-3}$ density profile that extends to $10^{16}$~cm (e.g., see Fig. \ref{fig:density}). Furthermore, the r1w4 model ($\dot M = 10^{-3}$~\mdot) cannot reproduce the spectral evolution of \sn{} because the IIn-like features are not sustained for a long enough timescale. Additionally, matching the observed $t_{\rm IIn}$ in \sn{} to the model grid produces only three consistent models: m1em2, r1w6, r1w6a, all with $\dot M = 10^{-2}$~\mdot. As shown in Figure \ref{fig:LC_colors}, the complete m1em2 and r1w6a model light curves provide a decent match to the multi-band \sn{} light curve at maximum light but cannot capture the steep rise to peak. This inconsistency is likely the result of the CSM density profile immediately beyond $R_{\star}$ and/or the limitations of the simulations at very early phases (e.g., $\delta t < 1$~day) e.g., light travel time effects which are considered in the radiation hydrodynamics simulations but are not included in the computation of the resulting spectrum.


\begin{table}[t!]
\begin{center}
\caption{Main parameters of \sn{} \label{tbl:params}}
\vskip0.1in
\begin{tabular}{lccc}
\hline
\hline
Host Galaxy &  &  NGC~3621 \\ 
Redshift &  &  0.002215\\  
Distance &  &  $7.2\pm 0.2$~Mpc\footnote{\cite{Saha06}}\\ 
Time of First Light (MJD) &  &  $60410.80 \pm 0.34$\\
$E(B-V)_{\textrm{MW}}$ &  &  0.07~mag\footnote{\cite{schlegel98,schlafly11}}\\
$E(B-V)_{\textrm{host}}$ &  &  $0.084\pm0.018$~mag\footnote{\cite{Stritzinger18}}\\
$M_{w2}^{\mathrm{peak}}$[$t_r$] &  &  $-18.7\pm0.07$~mag[$3.0\pm0.3$~d]\\
$M_{m2}^{\mathrm{peak}}$[$t_r$] &  &  $-18.7\pm0.07$~mag[$3.1\pm0.3$~d]\\
$M_{w1}^{\mathrm{peak}}$[$t_r$] &  &  $-18.4\pm0.07$~mag[$3.2\pm0.3$~d]\\
$M_{u}^{\mathrm{peak}}$[$t_r$] &  &  $-18.2\pm0.06$~mag[$3.5\pm0.3$~d]\\
$M_{g}^{\mathrm{peak}}$[$t_r$] &  &  $-18.1\pm0.06$~mag[$6.5\pm0.9$~d]\\
$M_{r}^{\mathrm{peak}}$[$t_r$] &  & $-17.8\pm0.14$~mag[$8.6\pm1.4$~d]\\
$M_{i}^{\mathrm{peak}}$[$t_r$] &  & $-17.7\pm0.09$~mag[$8.5\pm2.6$~d]\\
R$_{\rm CSM}$ &  &  $\sim 5 \times 10^{14}$~cm\\
M$_{\rm CSM}$ &  &  $(0.02-0.04)~\Msun$\\
$\dot{M}$\footnote{Mass-loss within $r<10^{15}$~cm} &  &  $10^{-2}~\Msun$~yr$^{-1}$\\
$v_w$ &  &  $50~\kms$\\
CSM Composition &  &  Solar Metallicity\footnote{Not varied in model grid}\\
Time of $\dot{M}$ &  &  $\sim$3~years pre-SN\\
\hline
\end{tabular}
\end{center}
\end{table}

\begin{figure*}
\centering
\subfigure{\includegraphics[width=0.49\textwidth]{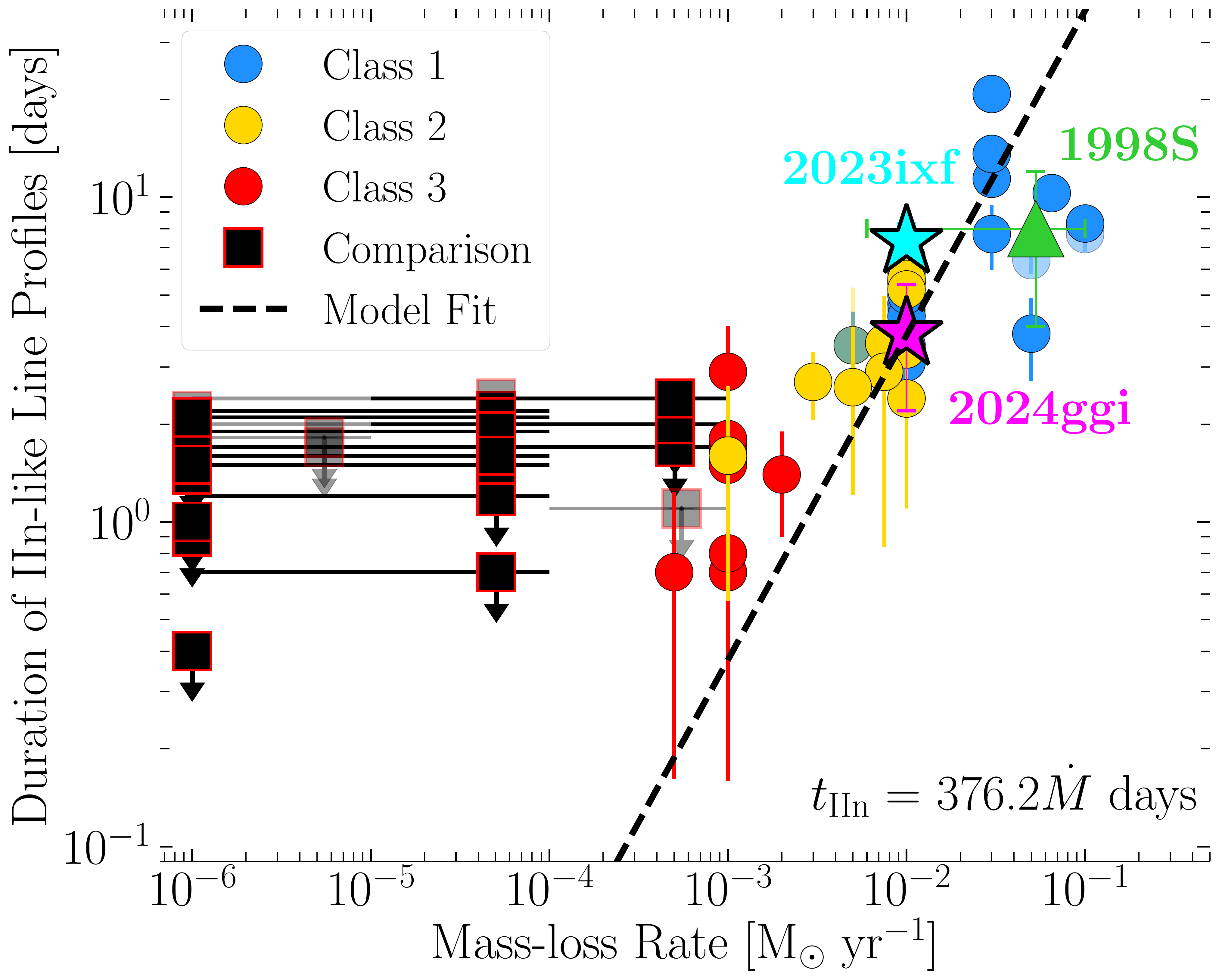}}
\subfigure{\includegraphics[width=0.49\textwidth]{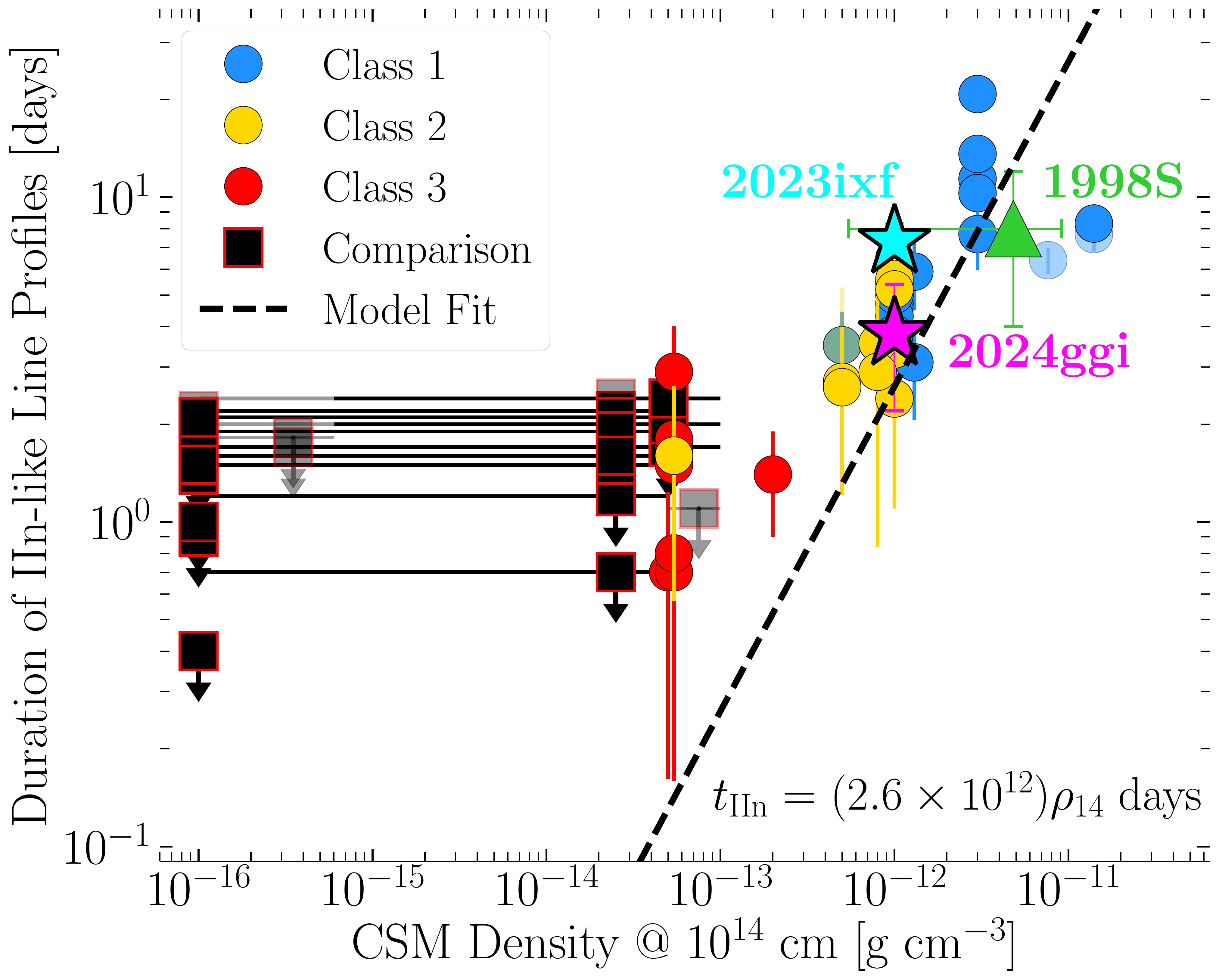}}
\caption{Duration of IIn-like features versus mass-loss rates {\it (left)} and CSM densities at $r = 10^{14}$~cm {\it (right)} derived from direct spectral matching for gold/silver- (blue, yellow and red circles) and comparison- (black squares) sample objects presented in \cite{wjg24}. Solid colored points represent the subsample of objects at $D>40$~Mpc. Best-matched mass-loss rate and CSM densities at $r = 10^{14}$~cm for \sn{} from spectral matching (\S\ref{subsec:modeling}) shown as magenta star. SNe~1998S and 2023ixf are shown for reference as a green triangle and cyan star, respectively.
\label{fig:mdot} }
\end{figure*}

\section{Discussion} \label{sec:discussion}


Observations of \sn{} for 2 weeks after first light have helped to constrain both the explosion dynamics/physics as well as progenitor activity prior to core-collapse. Specifically, the observed change in ionization present in the $\sim 0.8 - 1.5$~day spectra (Fig. \ref{fig:vels}) and the rise in temperature shown by the red-to-blue color evolution (Fig. \ref{fig:LC_colors}) suggests that we caught the SN as the shock broke out of the optically-thick envelope and into the extended, dense, and initially cold CSM. Rather than a prompt rise in temperature to $\sim10^5$~K, the CSM heats up to a lower temperature and becomes progressively more ionized (e.g., \ion{He}{i/ii}, \ion{C}{iii/iv}, or \ion{N}{iii/iv/v}) on a $\sim1-2$~day timescale as the radiative precursor, and then the radiation from the shock/ejecta, diffuses through the CSM. Furthermore, comparison of the spectral series, $t_{\rm IIn}$, peak luminosities and rise-times to a grid of \cmfgen\ models indicates that the progenitor of \sn{} was likely a RSG with a mass-loss rate of $\dot{M} \approx 10^{-2}~\Msun$~yr$^{-1}$, which created dense CSM extending to $r \approx 5\times10^{14}$~cm that contained a total mass of $M_{\rm CSM} \approx 0.02-0.04~\Msun$. This range of CSM masses is consistent with simulations of RSG mass-loss through energy injection within convective RSG envelopes \citep{Tsang22}, pre-SN outbursts \citep{Takei22}, and RSG ``boil-off'' \citep{Fuller24}. We also find that the observed light curve can be matched with standard explosion energy ($1.2\times 10^{51}$~erg) and that the IIn-like signatures in \sn{} can be modeled with a CSM composition that matches typical RSG surface abundances, which suggests that significant N or He enrichment to the CSM may not be required.

For a wind velocity of $\sim 50~\kms$, the proposed CSM extent translates to a period of enhanced mass-loss in the last $\sim$3~years years prior to core-collapse. Similarly, for the observed $t_{\rm IIn}$ and a shock velocity of $v_{\rm sh} = 12000~\kms$, the dense CSM would extend to $r \approx (4.3\pm1.7)\times 10^{14}$~cm and would indicate enhanced mass-loss in the final $\sim1.4-3.6$~years before first light, for a progenitor wind velocity of $50~\kms$. However, it should be noted that this timescale may not be accurate for enhanced wind scenarios (e.g., \citealt{davies22}) because the timescales for wind acceleration to a terminal velocity can take longer than the above pre-SN timescale. From the spectral evolution of \sn{} in its first two weeks, it is clear that a reduction in CSM density did occur at $\sim 3.8$~days because as the CSM optical depth to electron-scattering drops below unity, the Lorentzian wings of the IIn-like features disappear and absorption profiles from the fastest moving material become visible. However, the open question is whether the CSM detected in \sn{} represents the only high-density shell of CSM (i.e. only one phase of enhanced mass-loss) that extends to $r<5\times 10^{14}$~cm (e.g., r1w6 model density profile) or if \sn{} sustained enhanced mass-loss for several decades before explosion (e.g., m1em2 model density profile). As was done for SN~2023ixf, future multi-wavelength monitoring will be able to determine the CSM density at larger scales as the SN shock samples material that was liberated from the progenitor star years-to-decades before explosion (\citealt{Grefenstette23, Berger23, Chandra24, Panjkov23}, Nayana et al., in prep). Overall, both the confined high-density CSM shell and the extended low-density wind may have made the RSG progenitor star quite dust-obscured prior to explosion \citep{davies22}, similar to what was revealed by pre-explosion imaging of the SN~2023ixf RSG progenitor \citep{Kilpatrick23, Jencson23, Qin23, Soraisam23, VanDyk23, Niu23}.

As shown in Figures \ref{fig:max_lines} \& \ref{fig:mdot}, the observables and best-matched CSM properties of \sn{} reside naturally within the continuum of SNe~II, both with and without spectroscopic signatures of interaction with dense CSM. \sn{}'s best-matched $\dot M$ and CSM density at $10^{14}$~cm is lower than that inferred for more extreme Class 1 events such as SNe~1998S, 2017ahn, 2018zd, 2020tlf, 2020pni and 2023ixf, all of which likely have higher mass-loss rates, and/or dense CSM that extends to larger radii. Nonetheless, \sn{}'s evolution requires a progenitor mass-loss rate that is significantly higher than what is inferred for standard SNe~II as well as what is measured in local RSGs \citep{Beasor20}. Overall, this continues to point towards a phase of enhanced mass-loss in the final years before explosion for some significant fraction of RSGs. 

\section{Conclusions} \label{sec:conclusion}

In this paper we have presented UV/optical/NIR observations of the nearby SN~II, 2024ggi located in nearby spiral host galaxy NGC~3621 at $D=7.2$~Mpc. Below we summarize the primary observational findings of \sn{}:  

\begin{itemize}

\item The first optical spectrum of \sn{} shows prominent narrow emission lines of \ion{H}{i}, \ion{He}{i}, \ion{N}{iii} and \ion{C}{iii} that result from the photo-ionization of dense, optically thick CSM. Between $\delta t = 0.8 - 1.5$~days, \sn{} exhibits a detectable rise in temperature and ionization that manifests as the appearance of higher ionization species such as \ion{He}{ii}, \ion{N}{iv/v}, \ion{C}{iv}, and \ion{O}{v}. This spectral evolution is temporally consistent with a dramatic red-to-blue evolution in $w2-v$ colors and an increasing blackbody temperature. These phenomena suggest that \sn{} was observed during the initial precursor associated with shock breakout inside dense CSM. 

\item \sn{} displayed electron-scattering broadened profiles (i.e., IIn-like) that persist on a characteristic timescale of $t_{\rm IIn} = 3.8 \pm 1.6$~days, after which time a decrease in optical depth to electron scattering due to a reduction in CSM density allows for the detection of broad absorption profiles from the fastest H-rich SN ejecta.

\item Interaction of the \sn{} ejecta with dense, confined CSM yielded peak UV/optical absolute magnitudes (e.g., $M_{\rm w2} = -18.7$~mag, $M_{\rm g} = -18.1$~mag) that are consistent with other SNe~II with IIn-like features presented in sample studies (e.g., \citealt{wjg24}). 

\item Comparison of \sn{}'s spectral evolution, peak absolute magnitudes, rise-times and duration of IIn-like profiles to a grid of \cmfgen\ simulations suggests a CSM that has a composition typical of a solar-metallicity RSG, confined to $r < 5 \times 10^{14}$~cm, and that is formed from a progenitor mass-loss rate of $\dot{M} = 10^{-2}~\Msun$~yr$^{-1}$ (i.e., $\rho \approx 10^{-12}$~g~cm$^{-3}$ at $r = 10^{14}$~cm). Adopting a wind velocity of $v_w = 50~\kms$, this scenario corresponds to a period of enhanced mass-loss during the last $\sim$3~years years before core-collapse.

\item \sn{} is similar to SN~2023ixf in its rise in ionization within $\sim$days of first light and the high-ionization species present in its early-time spectra, both objects likely observed during shock breakout within dense CSM. However, \sn{} has a shorter timescale of IIn-like features ($\sim4$~days vs. $\sim7$~days) and a more compact CSM ($<5\times10^{14}$~cm vs $(0.5-1)\times 10^{15}$~cm) despite a similar mass-loss rate and CSM density at $10^{14}$~cm. 

\end{itemize}

Like SN~2023ixf, \sn{} represents a unique opportunity to study the long-term, multi-wavelength evolution of a SN~II that interact with dense, confined CSM in exquisite detail. Future studies and multi-wavelength observations will probe the nature and late-stage evolution of the progenitor star as well as uncover the properties of the large-scale circumstellar environment, both of which tracing an unconstrained period of RSG evolution before core-collapse.

\section{Acknowledgements} \label{Sec:ack}

Research at UC Berkeley is conducted on the territory of Huichin, the ancestral and unceded land of the Chochenyo speaking Ohlone people, the successors of the sovereign Verona Band of Alameda County. Shane 3-m observations were conducted on the land of the Ohlone (Costanoans), Tamyen and Muwekma Ohlone tribes.

W.J.-G.\ is supported by the National Science Foundation Graduate Research Fellowship Program under Grant No.~DGE-1842165. W.J.-G.\ acknowledges support through NASA grants in support of {\it Hubble Space Telescope} program GO-16075 and 16500. This research was supported in part by the National Science Foundation under Grant No. NSF PHY-1748958.  R.M.\ acknowledges support by the National Science Foundation under Award No. AST-2221789
and AST-2224255.  The Margutti team at UC Berkeley is partially funded by the Heising-Simons Foundation under grant \# 2018-0911 and \#2021-3248 (PI: Margutti).

MRD acknowledges support from the NSERC through grant RGPIN-2019-06186, the Canada Research Chairs Program, and the Dunlap Institute at the University of Toronto. Parts of this research were supported by the Australian Research Council Discovery Early Career Researcher Award (DECRA) through project number DE230101069. Y.-C.P.\ is supported by the National Science and Technology Council (NSTC grant 109-2112-M-008-031-MY3. CG is supported by a VILLUM FONDEN Young Investigator Grant (project number 25501). D.M.\ acknowledges NSF support from grants PHY-2209451 and AST-2206532. A.S.\ acknowledges the financial support from CNPq (402577/2022-1). G.D.\ is supported by the H2020 European Research Council grant no.758638. CRB acknowledges the financial support from CNPq (316072/2021-4) and from FAPERJ (grants 201.456/2022 and 210.330/2022) and the FINEP contract 01.22.0505.00 (ref. 1891/22). This research is based on observations made with the Thai Robotic Telescope under program ID TRTC11B006, which is operated by the National Astronomical Research Institute of Thailand (Public Organization). Part of the data were obtained with the REM telescope, located in Chile and operated by the d'REM team for INAF. This work is supported by the National Science Foundation under Cooperative Agreement PHY-2019786 (The NSF AI Institute for Artificial Intelligence and Fundamental Interactions, http://iaifi.org/).

The UCSC team is supported in part by NASA grants NNG17PX03C and 80NSSC22K1518, NSF grant AST--1815935, and by a fellowship from the David and Lucile Packard Foundation to R.J.F.

YSE-PZ \citep{Coulter23} was developed by the UC Santa Cruz Transients Team with support from NASA grants NNG17PX03C, 80NSSC19K1386, and 80NSSC20K0953; NSF grants AST-1518052, AST-1815935, and AST-1911206; the Gordon \& Betty Moore Foundation; the Heising-Simons Foundation; a fellowship from the David and Lucile Packard Foundation to R.J.F.; Gordon and Betty Moore Foundation postdoctoral fellowships and a NASA Einstein fellowship, as administered through the NASA Hubble Fellowship program and grant HST-HF2-51462.001, to D.O.J.; and a National Science Foundation Graduate Research Fellowship, administered through grant No.\ DGE-1339067, to D.A.C.

A major upgrade of the Kast spectrograph on the Shane 3~m telescope at Lick Observatory, led by Brad Holden, was made possible through generous gifts from the Heising-Simons Foundation, William and Marina Kast, and the University of California Observatories. Research at Lick Observatory is partially supported by a generous gift from Google.

This work has made use of data from the Asteroid Terrestrial-impact Last Alert System (ATLAS) project. The Asteroid Terrestrial-impact Last Alert System (ATLAS) project is primarily funded to search for near earth asteroids through NASA grants NN12AR55G, 80NSSC18K0284, and 80NSSC18K1575; byproducts of the NEO search include images and catalogs from the survey area. This work was partially funded by Kepler/K2 grant J1944/80NSSC19K0112 and HST GO-15889, and STFC grants ST/T000198/1 and ST/S006109/1. The ATLAS science products have been made possible through the contributions of the University of Hawaii Institute for Astronomy, the Queen’s University Belfast, the Space Telescope Science Institute, the South African Astronomical Observatory, and The Millennium Institute of Astrophysics (MAS), Chile.

This research was supported by the Munich Institute for Astro-, Particle and BioPhysics (MIAPbP) which is funded by the Deutsche Forschungsgemeinschaft (DFG, German Research Foundation) under Germany's Excellence Strategy – EXC-2094 – 390783311.

IRAF is distributed by NOAO, which is operated by AURA, Inc., under cooperative agreement with the National Science Foundation (NSF).

\facilities{ATLAS, {\it Swift} UVOT, Gemini-South Observatory (GMOS), SOAR (Goodman/TSpec), Shane Telescope (Kast), Thailand Robotic Telescope, Las Cumbres Observatory, Lulin Telescope, Cerro Tololo Inter-american Observatory, Rapid Eye Mount Telescope}

\software{IRAF (Tody 1986, Tody 1993), photpipe \citep{Rest+05}, DoPhot \citep{Schechter+93}, HOTPANTS \citep{becker15}, YSE-PZ \citep{Coulter22, Coulter23}, \cmfgen\ \citep{hillier12, D15_2n}, \heracles\ \citep{gonzalez_heracles_07,vaytet_mg_11,D15_2n}, HEAsoft (v6.33; HEASARC 2014), DRAGONS \citep{Labrie23} }

\bibliographystyle{aasjournal} 
\bibliography{references} 


\clearpage
\appendix

Here we present a log of optical spectroscopic observations of \sn{} in Table \ref{tab:spec_table}.

\renewcommand\thetable{A\arabic{table}} 
\setcounter{table}{0}

\begin{deluxetable*}{cccccc}[h!]
\tablecaption{Optical Spectroscopy of SN~2024ggi \label{tab:spec_table}}
\tablecolumns{5}
\tablewidth{0.45\textwidth}
\tablehead{
\colhead{UT Date} & \colhead{MJD} &
\colhead{Phase\tablenotemark{a}} &
\colhead{Telescope} & \colhead{Instrument} & \colhead{Wavelength Range}\\
\colhead{} & \colhead{} & \colhead{(days)} & \colhead{} & \colhead{} & \colhead{(\AA)}
}
\startdata
2024-04-12T06:10:08.92 & 60412.26 & 1.46 & Shane & Kast & 3254--10495 \\
2024-04-13T00:00:05.88 & 60413.00 & 2.20 & SOAR & TripleSpec & 8243.4--24667.1 \\
2024-04-13T00:41:21.57 & 60413.03 & 2.22 & SOAR & Goodman & 3816--7030 \\
2024-04-16 06:21:35.01 & 60416.26 & 5.46 & Shane & Kast & 3603--10293 \\
2024-04-16T06:34:06.69 & 60416.27 & 5.47 & Shane & Kast & 5800--7000 \\
2024-04-18T05:49:43.62 & 60418.24 & 7.44 & Shane & Kast & 3253--10494 \\
2024-04-19T23:41:17.00 & 60419.99 & 9.19 & Gemini-S & GMOS & 4000--7580 \\
2024-04-20T01:32:39.64 & 60420.06 & 9.26 & SOAR & TripleSpec & 8243.0--24667.5 \\
2024-04-20T02:19:26.89 & 60420.10 & 9.30 & SOAR & Goodman & 3854--8986 \\
\enddata
\tablenotetext{a}{Relative to first light (MJD 60410.80)}
\end{deluxetable*}

\end{document}